\documentclass[aps,prc,twocolumn,superscriptaddress]{revtex4}

\usepackage{amsmath}  
\usepackage{amsfonts}
\usepackage{amssymb}
\usepackage{natbib} % Use the natbib bibliography and citation package
\usepackage{setspace}
\usepackage{graphicx} 
\usepackage{xspace}
\usepackage{cancel}
\usepackage{url}
\usepackage{color}
\usepackage{float}

\newcommand{\ie}{\text{i.e.}\xspace}

\newcommand{\dx}[1]{\hspace{-0.4em}\ensuremath{\mathrm{d}#1}\,}

\newcommand{\eqn}[1]{Eq.~(\ref{#1})}
\newcommand{\fig}[1]{Fig.~\ref{#1}}
\newcommand{\tab}[1]{Table~\ref{#1}}

\newcommand{\sect}[1]{Sec~\ref{#1}}

\newcommand{\be}{\begin{equation}}
\newcommand{\ee}{\end{equation}}
\newcommand{\bge}{\begin{equation}}
\newcommand{\ene}{\end{equation}}
\newcommand{\bea}{\begin{eqnarray}}
\newcommand{\eea}{\end{eqnarray}}
\newcommand{\bg}{\begin{eqnarray}}
\newcommand{\en}{\end{eqnarray}}

\usepackage[normalem]{ulem}
\usepackage{color}

\def\bred{\bf\boldmath\color{red}}

\begin{document}

\title{
  \vspace{-50mm}
  \begin{flushright}
    {\bred LFTC-22-1/69}
  \end{flushright}
  \vspace{10mm}
 $\Upsilon$ and $\eta_b$ nuclear bound states}

\author{J.~J.~Cobos-Mart\'{\i}nez}
\affiliation{Departamento de F\'isica, Universidad de Sonora, Boulevard
Luis Encinas J. y Rosales, Colonia Centro, Hermosillo, Sonora 83000, M\'exico}
\author{G.~N.~Zeminiani }
\author{K.~Tsushima}
\affiliation{Laborat\'orio de F\'isica Te\'orica e Computacional, 
Universidade Cidade de S\~ao Paulo (UNICID), 
01506-000, S\~ao Paulo, SP, Brazil}

\date{\today}

\begin{abstract}
   $\Upsilon$ and $\eta_b$ nuclear bound state energies are calculated for 
various nuclei neglecting any possible effects of the widths. 
Essential input for the calculations, namely the medium-modified $B$ and $B^{*}$ meson masses,  as 
well as the density distributions in nuclei, are calculated 
within the quark-meson coupling (QMC) model. 
The attractive potentials for the $\Upsilon$ and $\eta_b$ mesons in nuclei
are calculated from the mass shifts of these mesons in nuclear matter in the local
density approximation. These potentials originate from the in-medium enhanced 
$B\overline{B}$ and $BB^{*}$ loops in their respective self energy.
After an extensive analysis we conclude that our results suggest that the $\Upsilon$ and 
$\eta_b$ mesons  should form bound states with all the nuclei considered.
\end{abstract}

%\pacs{pacs here}
%\keywords{keywords here}

\maketitle

\date{\today}

\section{Introduction}

Quantum chromodynamics (QCD) is the accepted theory of the strong interactions at the
fundamental level. However, a quantitative understanding of the strong force and strongly interacting
matter from the underlying theory is still limited.
The study of the interactions between heavy quarkonia and atomic nuclei is an important tool to 
gain an understanding of the  strongly interacting matter 
properties in vacuum and extreme conditions 
of temperature and density based on QCD.

Since  heavy quarkonium and nucleons do not share light ($u$, $d$) quarks (the OZI rule suppresses the
interactions mediated by the exchange of mesons made of only light quarks), heavy quarkonium 
interacts with nucleon primarily via gluons, and therefore the production of heavy quarkonium in a
nuclear  medium can be of great relevance to explore the role played by gluons. 
If such states are indeed found experimentally to be bound to nuclei, it is therefore important to search 
for other sources of  attraction which could lead to the binding of  heavy quarkonium to nuclei.  
The binding of heavy quarkonium to nuclei may give an evidence that the 
masses of these heavy mesons decrease in a nuclear medium.

Since the early work of Brodsky~\cite{Brodsky:1989jd} that charmonium states may be bound 
to nuclei, a large amount of research, looking for alternatives to the light meson exchange mechanism,
has accumulated over the years to investigate the possible existence  of such exotic 
states~\cite{Hosaka:2016ypm,Krein:2016fqh,Metag:2017yuh,Krein:2017usp,Ko:2000jx,Krein:2010vp,
Tsushima:2011kh,Tsushima:2011fg,Krein:2013rha,Klingl:1998sr,Hayashigaki:1998ey,Kumar:2010hs,
Belyaev:2006vn,Yokota:2013sfa,Peskin:1979va,Kharzeev:1995ij,Kaidalov:1992hd,Luke:1992tm,
deTeramond:1997ny,Brodsky:1997gh,
Sibirtsev:2005ex,Voloshin:2007dx,TarrusCastella:2018php}.
In addition to these, lattice QCD simulations for charmonium-nucleon interaction in free space were
performed in the last 
decade~\cite{Yokokawa:2006td,Liu:2008rza,Kawanai:2010ev,Kawanai:2010ru,Skerbis:2018lew}. 
Furthermore, more recently, studies for the binding of charmonia with nuclear
 matter and finite nuclei,  as well as light mesons and baryons, were performed in lattice QCD 
 simulations~\cite{Beane:2014sda,Alberti:2016dru}, albeit with unphysically heavy pion masses.

On the experimental side, the 12 GeV upgrade at the Jefferson Lab has made it possible to produce
low-momentum heavy-quarkonia in an atomic nucleus. Recently~\cite{Ali:2019lzf}, a photon beam 
was  used to produce  a $J/\Psi$ meson near-threshold, which was identified by the decay into an 
electron-positron pair. Furthermore, with the construction of the FAIR facility in Germany, heavy and 
heavy-light mesons will be produced copiously by the annihilation of antiprotons 
on nuclei~\cite{Durante:2019hzd}.  
Experimental studies on $\eta_c$ production in heavy ion collisions at 
the LHC  were  performed in Refs.~\cite{Aaij:2019gsn,Tichouk:2020dut,Tichouk:2020zhh,Goncalves:2018yxc,Klein:2018ypk}.  However, nearly no experiments have yet been aimed to produce the $\eta_c$ 
at lower energies and its binding to nuclei, perhaps hinting at the difficulty to produce and detect such states. 
In the case of bottomonium, studies were made for $\Upsilon$ photoproduction at the Electron-Ion Collider~\cite{Xu:2020uaa,Gryniuk:2020mlh},  $\Upsilon$ production in $p$Pb 
collisions~\cite{Aaij:2018scz},  and $\Upsilon(nl)$ (excited state) decay into $ B^{(*)} \bar B^{(*)}$~\cite{Liang:2019geg}. 
With studies like these on heavy quarkonium and future planned ones, we will improve our understanding 
of the strong force and strongly interacting matter.

Returning to the phenonenological studies, the interactions frequently considered between the 
heavy quarkonium and  the nuclear medium are the so called QCD van der Waals (multigluon exchange) 
interactions~\cite{Kaidalov:1992hd,Luke:1992tm, deTeramond:1997ny,Brodsky:1997gh,Ko:2000jx,Sibirtsev:2005ex, Voloshin:2007dx, TarrusCastella:2018php}. One might think that this must 
be the case, since heavy quarkonium has no light quarks,  whereas 
the nuclear medium is composed of light  quarks, and thus the  exchange of mesons composed of
light quarks do not occur at the lowest order. 

However, another possible mechanism, which we consider in this paper, for the heavy quarkonium
to interact with the nuclear medium is through the excitation of the intermediate state hadrons 
which do contain light quarks ($B$ and $B^*$ in this work).
There is a great amount of evidence that the internal structure of hadrons changes in medium
and this must be taken into account when addressing, for example, charmonium in nuclei.
For instance, Refs.~\cite{Krein:2010vp,Tsushima:2011kh} have shown that the effect of the nuclear 
mean fields on  the intermediate $DD$ state is crucial when considering the 
$J/\Psi$ interactions with atomic nuclei. The modifications driven by the strong nuclear mean
fields on the $D$ mesons’ light-quark component enhanced the self-energy such that it provides
attraction to the  $J/\Psi$.
Furthermore, only recently the in-medium properties of $\eta_c$ meson were renewed 
theoretically~\cite{Cobos-Martinez:2020ynh} using this mechanism.

\begin{figure}[htb]
\centering 
  \includegraphics[scale=0.9]{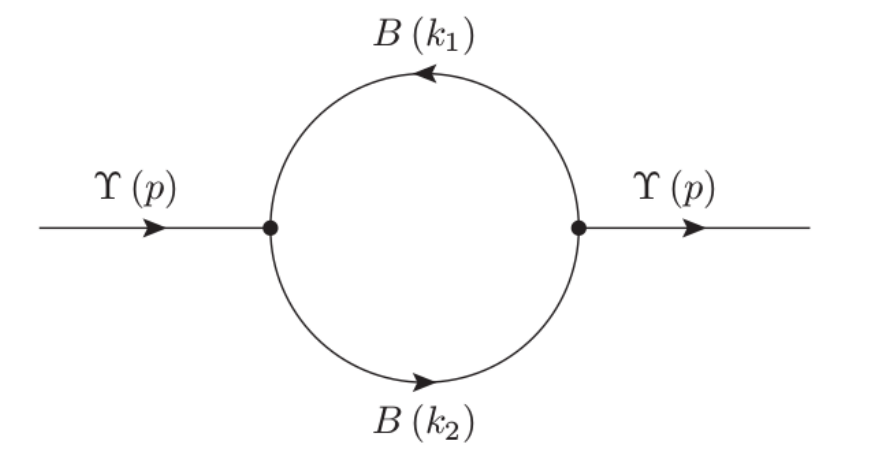}
 \caption{BB meson loop contribution for the $\Upsilon$ self-energy.}
 \label{fig1}
\end{figure}

In a recent paper~\cite{Zeminiani:2020aho},  we estimated the mass shifts of the $\Upsilon$ and 
$\eta_b$ mesons by considering the excitations of intermediate state hadrons with light quarks in 
their self-energy. 
The estimates were made using an SU(5) effective Lagrangian density which contains both the 
$\Upsilon$ and $\eta_b$ mesons with one universal coupling constant, and an anomalous coupling  
that respects SU(5)  symmetry in the coupling constant.  After expansion of the SU(5) effective
 Lagrangian with minimal substitutions, we obtained the interaction Lagrangians for calculating 
 the the $BB$, and $B^*B^*$ meson loops contributions to the  $\Upsilon$  and $\eta_b$ self energies.
As an example we show in Fig.~\ref{fig1} the $BB$ meson loop contribution for the $\Upsilon$ 
self-energy.

%%%%%%%%%%%%%%%%%%%%%%%%%%%%%%%%%%%%%%%%%%%%%%%%%%%%%%%%%%%%%%
%%%%%%%%%%%%%%%%%%%%%%%%%%%%%%%%%%%%%%%%%%%%%%%%%%%%%%%%%%%%%

For the study of the heavy quarkonium ($\Upsilon$ and $\eta_b$) interaction with the nuclear medium
 through the excitation of the intermediate state hadrons we need to have  knowledge on the in-medium
  properties  of the $B$ and  $B^{*}$ mesons, in particular their medium-modified masses.
For this we used the quark-meson coupling (QMC) model~\cite{Guichon:1987jp},  
which has been successfully applied for various studies in nuclear matter and 
nuclei~\cite{Krein:2017usp,Krein:2010vp,Tsushima:2002cc,Tsushima:1997df,Guichon:1995ue, 
Saito:1996sf,Tsushima:2019wmq,Saito:2005rv,Guichon:1989tx,Stone:2016qmi}.

In Ref.~\cite{Zeminiani:2020aho} we did an study of the $BB, BB^*$ and $B^*B^*$ meson
loop contributions to the $\Upsilon$ self-energy in nuclear matter neglecting for the moment
any possible  imaginary part.
After a detailed analysis, our predictions for the $\Upsilon$ and $\eta_b$ mass shifts
were given by including only the lowest order $BB$ meson loop contribution for 
the $\Upsilon$,  and only the $BB^*$ meson loop contribution for the $\eta_b$,  
where the in-medium masses of the $B$ and $B^*$ mesons were calculated by the QMC model.
We note that the in-medium $B^*$ meson mass was calculated for the first 
time in Ref.~\cite{Zeminiani:2020aho}. 
In this work, we apply the mechanism  described above,  by first extending our results in nuclear 
matter to finite nuclei, and then considering the interactions between the $\Upsilon$ and 
$\eta_b$ mesons and a wide mass range of atomic nuclei.

This article is organized as follows.  In Sec.~\ref{sec:nuclmatt} we summarize the computational 
procedure used and discuss our results for the mass shifts of the $\Upsilon$ and $\eta_b$ mesons in 
nuclear matter. These results indicate that nuclear medium provides attraction to these mesons 
and therefore in Sec.~\ref{sec:finitenuclei} we consider the nuclear bound states for the $\Upsilon$ 
and $\eta_b$ mesons when these mesons are produced nearly at rest inside a nucleus.
Finally, in Sec.~\ref{sec:summary} we give a summary and conclusions.

\section{\label{sec:nuclmatt} Mass shifts in nuclear matter}

In this section we summarize the results obtained for the mass shifts of the $\Upsilon$ and
$\eta_b$ mesons in nuclear matter. The details of this  analysis can be found in 
Ref.~\cite{Zeminiani:2020aho}.

\subsection{$\Upsilon$ mass shift}

The $\Upsilon$ mass shift in nuclear matter originates from the modifications of the 
$BB$, $BB^{*}$, and $B^{*}B^{*}$ meson loops contributions to the $\Upsilon$ self-energy, relative to those in free space; see, for example, Fig.~\ref{fig1}.  The self-energy is calculated using 
effective $SU(5)$-flavor symmetric Lagrangians at the hadronic 
level~\cite{Zeminiani:2020aho,Lin:2000ke} for the interaction vertices $\Upsilon B B$, $\Upsilon 
B^{*}B^{*}$, and $\Upsilon B B^{*}$ neglecting any possible imaginary part.
In  Ref.~\cite{Zeminiani:2020aho} we have made an extensive analysis of these contributions to
the $\Upsilon$ self-energy and have found that, for example, the $B^{*}B^{*}$ loop gives an
unexpectedly large contribution. For this reason, and to be consistent with the $\eta_b$ case studied
below, we have decided to be conservative and consider only the $BB$ loop contribution
to the $\Upsilon$ self-energy, leaving for the future a full study. The interaction Lagrangian for the
$\Upsilon B B$ vertex is given by
\begin{equation}
\label{eqn:LUBB}
{\cal L}_{\Upsilon BB} 
= i g_{\Upsilon BB}\Upsilon^{\mu} 
\left[\overline{B} \partial_{\mu}B - \left(\partial_{\mu}\overline{B}\right)B\right],
\end{equation}
where the following convention is adopted for the isospin doublets of the $B$ mesons
\begin{align*}
B&=\begin{pmatrix}
        B^{+}\\
        B^{0}
       \end{pmatrix}, & \overline{B}=\begin{pmatrix}
       B^{-} & \overline{B^{0}} \end{pmatrix} .  
\end{align*}
The coupling constant $g_{\Upsilon BB}$ for the vertex $\Upsilon B B$ is calculated from the 
experimental data for $\Gamma(\Upsilon\to e^{+}e^{-})$ using the vector meson dominance model. This
gives $g_{\Upsilon BB}=13.2$; see Refs.~\cite{Zeminiani:2020aho,
Lin:2000ke} and references therein for details. A similar approach was taken in 
Refs.~\cite{Lin:1999ad,Krein:2010vp} to determine the coupling constant $g_{J/\Psi DD}=7.64$ for 
the vertex  $J/\Psi B B$.

Including only the $BB$ loop, \eqn{eqn:LUBB}, the $\Upsilon$ self-energy $\Sigma_{\Upsilon}$ is 
given by
\begin{equation}
    \label{eqn:upsilon_se}
    \Sigma_{\Upsilon}(k^{2})= -\frac{g_{\Upsilon B B}^{2}}{3\pi^{2}}\int_{0}^{\infty}
    \dx{q}\textbf{q}^2\,I(\textbf{q}^2)
\end{equation}
for an $\Upsilon$ at rest, where
\begin{equation}
I(\textbf{q}^2)=\frac{1}{\omega_B}\left(\frac{\textbf{q}^2}
{\omega_B-m_{\Upsilon}^2/4} \right)  
\end{equation}
and $\omega_B = \left(\textbf{q}^{2} + m^{2}_{B}\right)^{1/2}$.
The integral in \eqn{eqn:upsilon_se} is divergent and therefore needs 
to be regularized. To do this, we  introduce into the integrand of \eqn{eqn:upsilon_se}  a phenomenological vertex form factor 
\begin{equation}
\label{eqn:upsilon_ff}
u_{B}(\textbf{q}^{2}) = \left(\frac{\Lambda^{2}_{B} + m^{2}_{\Upsilon}}
{\Lambda^{2}_{B} + 4\omega^{2}_{B}\left(\textbf{q}^{2}\right)}\right)^{2},
\end{equation}
with cutoff parameter $\Lambda_{B}$~\cite{Krein:2010vp, Tsushima:2011kh, Tsushima:2011fg, Krein:2013rha, Cobos-Martinez:2017vtr, Cobos-Martinez:2017woo, Cobos-Martinez:2017onm,
  Cobos-Martinez:2017fch, Cobos-Martinez:2019kln}, for each $\Upsilon B B$ vertex.  In a
 later section we will discuss the non-neglegible role played by this form factor  and the cutoff
  parameter $\Lambda_{B}$.
For the moment, we point out that form factors are necessary to take into account the finite size of the mesons participating in the vertices, while the cutoff  $\Lambda_B$,
which is an unknown input to our calculation, may be associated with energies needed to probe the internal structure of the mesons; see Ref.~\cite{Zeminiani:2020aho} for a more extensive discussion.
Thus, in order to reasonably include these effects, and to  quantify
the sensitivity of our results to its value, we vary $\Lambda_B$ over the interval 2000-6000 MeV.
 
 %%%%%%%%%%%%%%%%%%
\begin{figure}[htb]
\vspace{4ex}
\centering
 \includegraphics[scale=0.33]{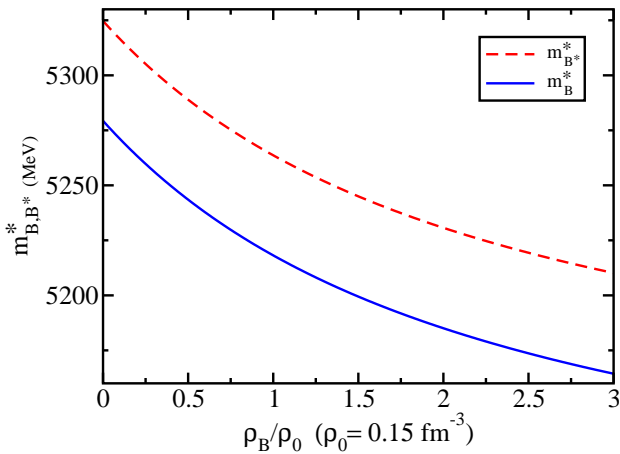}
 \caption{$B$ and $B^{*}$ meson effective Lorentz-scalar masses  
 in symmetric nuclear matter versus baryon density.}
 \label{fig:BBs_nm}
\end{figure}
%%%%%%%%%%%%%%%%%

%%%%%%%%%%%%%%%%%
\begin{figure}[htb]%
\vspace{4ex}
\centering
\includegraphics[scale=0.33]{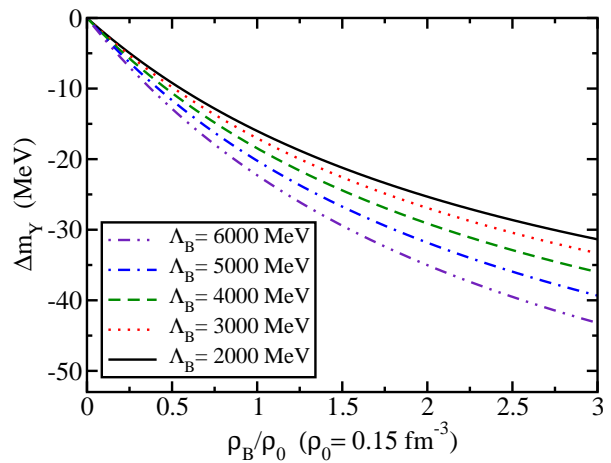}%     
\caption{$\Upsilon$ mass shift in nuclear matter as a function of the nuclear matter densidy $\rho_B$}%
\label{fig:upsilon_ms}%
\end{figure}
%%%%%%%%%%%%%%%%

The $\Upsilon$ mass shift in nuclear matter $\Delta m_{\Upsilon}$ is computed from the difference 
between its in-medium mass, $m_{\Upsilon}^{*}$, and its value in vacuum, $m_\Upsilon$, namely
\begin{equation}
\label{eqn:upsilon_ms}
\Delta m_{\Upsilon}= m_{\Upsilon}^{*} - m_\Upsilon,
\end{equation}
where these masses are computed  self-consistently from
\begin{equation}
\label{eqn:upsilon_mass}
    m_{\Upsilon}^{2}= (m_{\Upsilon}^{0})^{2} + \Sigma_{\Upsilon}(k^{2}=m_{\Upsilon}^{2}),
\end{equation}
where $m_{\Upsilon}^{0}$ is the bare $\Upsilon$ mass and the $\Upsilon$ self-energy
$\Sigma_{\Upsilon}(k^{2})$ is given in \eqn{eqn:upsilon_se}. 
The $\Lambda_{B}$-dependent $\Upsilon$ bare mass,  $m_{\Upsilon}^{0}$, is fixed such that
we reproduce the physical $\Upsilon$ mass, namely $m_{\Upsilon}=9640$ MeV.

The in-medium $\Upsilon$ mass is obtained by solving \eqn{eqn:upsilon_mass} with the self-energy
calculated with the  medium-modified $B$ mass, which was calculated in Ref.~\cite{Zeminiani:2020aho},
together with that for the $B^{*}$ meson, using the quark-meson coupling model (QMC) as a 
function of the nuclear matter density $\rho_B$. The results obtained in this way are presented in 
\fig{fig:BBs_nm} and show
that the QMC model gives a similar downward mass shift for the $B$ and $B^{*}$ in symmetric 
nuclear matter.
For example, at the saturation density $\rho_0=0.15\,\text{fm}^{-3}$, the mass shifts
for the $B$ and $B^*$ mesons are respectively,  $(m^*_B - m_B)=-61$ MeV and 
$(m^*_{B^*}-m_{B^*})=-61$ MeV, where  
the difference  in their mass shift values appears in the next digit.
The values for the masses in vacuum for the $B$ and $B^{*}$ mesons used are $m_B= 5279$ MeV and
$m_{B^{*}}=5325$ MeV, respectively.

The nuclear density dependence of the $\Upsilon$ mass is driven by the  intermediate $BB$ state
 interactions with the nuclear medium, where the effective scalar and vector meson mean fields 
 couple to the  light $u$ and $d$ quarks in the bottom mesons.  In \fig{fig:upsilon_ms} we show the
  results  for  the $\Upsilon$ mass shift as a function of the nuclear matter density $\rho_B$, for five 
  values of the cutoff parameter $\Lambda_B$. As can be seen from  Figs.~\ref{fig:BBs_nm} and 
  \ref{fig:upsilon_ms}, a decreasing  $B$ meson mass in-medium 
  induces a negative mass shift for the 
  $\Upsilon$. This happens because a  decrease of the $B$ meson mass enhances 
 the $BB$ meson loop contribution in nuclear matter relative to that in vacuum.
  Expectedly, the mass shift of the $\Upsilon$  is dependent on the value of the cutoff mass 
  $\Lambda_B$ used, being larger for larger $\Lambda_B$; see  Ref.~\cite{Zeminiani:2020aho} for 
  further details. 
  For example, for the values of the cutoff shown in \fig{fig:upsilon_ms}, the $\Upsilon$
  mass shift varies from -16 to -22 MeV, at  $\rho_B=\rho_0$.  
  %We will see later that a negative mass shift
  %for the $\Upsilon$ gives rise to an attractive potential between this meson and nuclei, 
  %opening the possibility to study the binding of $\Upsilon$ mesons to nuclei.

\subsection{$\eta_b$ mass shift}

%%%%%%%%%%%%%%%%%
\begin{figure}[htb]%
\vspace{4ex}
\centering
\includegraphics[scale=0.33]{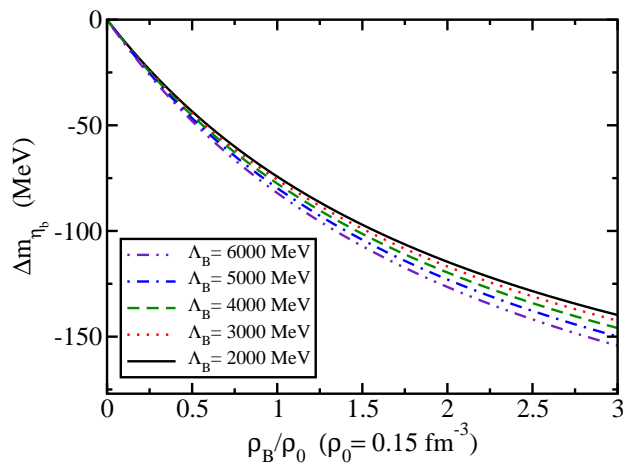}%     
\caption{$\eta_b$ mass shift in nuclear matter as a function of the nuclear matter densidy $\rho_B$.}%
\label{fig:etab_ms}%
\end{figure}
%%%%%%%%%%%%%%%%

For the calculation of the $\eta_b$ mass shift in nuclear matter, we proceed similarly to the
$\Upsilon$ case and take into account only the $BB^*$ loop contribution to the $\eta_b$ self-energy.
In Ref.~\cite{Zeminiani:2020aho} we have also studied the inclusion of the $\eta_b B^*B^*$
interaction in the $\eta_b$ self-energy and found that its contribution to the mass shift is essentially
negligible. 
Thus, in order to be consistent with the $\Upsilon$ case  above, \ie in both cases we consider only 
the minimal contribution, here we only give results for the $BB^*$ loop in the $\eta_b$ self-energy.
The effective Lagrangian for the $\eta_b BB^*$ interaction is
\begin{eqnarray}
\label{eqn:LetabBBast}
{\cal L}_{\eta_b BB^*} 
&=& i g_{\eta_b BB^*}
\left[ (\partial^\mu \eta_b) 
\left( \overline{B}^*_\mu B - \overline{B} B^*_\mu \right)\right. \nonumber  \\
&-& \left. \eta_b 
\left( \overline{B}^*_\mu (\partial^\mu B) - (\partial^\mu \overline{B}) B^*_\mu \right)
\right ], 
\end{eqnarray}
where $g_{\eta_b BB^*}$ is the coupling constant for the $\eta_b BB^*$ vertex. 
We will use its value in the  SU(5) scheme~\cite{Zeminiani:2020aho}, namely
\begin{equation}
\label{eqn:su5}
g_{\eta_b BB^*} = g_{\Upsilon BB} = g_{\Upsilon B^*B^*} = \frac{5g}{4\sqrt{10}}.
\end{equation}
Using \eqn{eqn:LetabBBast}, the $\eta_b$ self-energy for an $\eta_b$ at rest  is given 
by~\cite{Cobos-Martinez:2020ynh}
\begin{equation}
\label{eqn:etab_se}
\Sigma_{\eta_b} 
= \frac{8 g_{\eta_b B B^*}^{2}}{\pi^{2}}\int_{0}^{\infty}
    \dx{q} \textbf{q}^{2} K(\textbf{q}^{2}),  
\end{equation}
where 
\begin{eqnarray}
\hspace{-5mm}K(\textbf{q}^{2})
&=& \left. \frac{m_{\eta_b}^{2}(-1+q_0^2/m_{B^{*}}^{2})}
{(q_0^2-\omega_{B^{*}}^2) 
(q_{0}-m_{\eta_b}-\omega_{B})}\right|_{q_{0}=m_{\eta_b}- \omega_B} 
\nonumber \\
&+& \left. \frac{m_{\eta_b}^{2}(-1+q_0^2/m_{B^{*}}^{2})}
{(q_{0}-\omega_{B^{*}})\left((q_{0}-m_{\eta_b})^2-\omega_{B}^2\right) 
}\right|_{q_{0}=-\omega_{B^*}},  \nonumber \\
\label{IBBs}
\end{eqnarray}
and $\omega_{B^*}=(\textbf{q}^{2}+m_{B^*}^{2})^{1/2}$.

The mass of the $\eta_b$ meson, in vacuum and in nuclear matter, is computed similarly to the 
$\Upsilon$ case.  First, we introduce form factors, as in \eqn{eqn:upsilon_ff}, 
into each $\eta_b BB^*$ vertex, with $\Lambda_B=\Lambda_{B^{*}}$, in order to regularize the 
divergent integral in the self-energy, \eqn{eqn:etab_se}. Second,  we fix the value of the $\eta_b$ 
bare mass using the physical (vacuum) mass of the $\eta_b$,  namely $m_{\eta_b}=9399$ MeV, using 
\eqn{eqn:upsilon_mass} appropriately written  for the $\eta_b$ case. 
 Then, for the calculation of the $\eta_b$ mass shift in nuclear matter,  the self-energy 
 $\Sigma_{\eta_b} $ is computed using the medium-modified $B$ and $B^{*}$ masses calculated in
 the QMC model and shown in \fig{fig:BBs_nm}. The results for the $\eta_b$ mass shift in nuclear
  matter are shown in \fig{fig:etab_ms} as a function of the nuclear matter density $\rho_B$.
  Note that we use the same range of values for the cutoff mass $\Lambda_B$ as for the $\Upsilon$.
As can be seen from \fig{fig:etab_ms}, the mass of the $\eta_b$ is shifted downwards in nuclear
matter for all values of the cutoff $\Lambda_B$, similarly to the $\Upsilon$. 
For example, at  the normal density of nuclear matter $\rho_0$, the mass shift varies from 
-75 MeV to -82 MeV when the cutoff varies from $\Lambda_B=2000$ MeV to $\Lambda_B=6000$ MeV.
Similarly to the $\Upsilon$ mass shift, the dependence of the $\eta_b$ mass shift on the
values of the cutoff is small, for example, just -7 MeV when the cutoff is increased by a factor of 3
at $\rho_B=\rho_0$.

\subsection{\label{sec: mass shift discussion} Discussion of  the $\Upsilon$ and $\eta_b$ mass shifts results}

Surprisingly,  the  mass shift for the $\eta_b$ is larger than that of the $\Upsilon$ for the same
range of cutoff values explored; see Figs.~\ref{fig:upsilon_ms} and \ref{fig:etab_ms}. A similar difference 
in mass shifts was observed for the $J/\Psi$ and $\eta_c$ mesons 
in Refs.~\cite{Krein:2010vp,Cobos-Martinez:2020ynh}, using the corresponding Lagrangians in the $SU(4)$ flavor sector. 
As demonstrated in  Refs.~\cite{Zeminiani:2020aho,Cobos-Martinez:2020ynh},
these differences in the mass shifts for the $\eta_b$
and $\Upsilon$ are probably due to the following reasons: ({\bf a}) a badly broken 
$SU(5)$ symmetry such that the couplings
$g_{\eta_b BB^*}$ and $g_{\Upsilon BB}$ are very different, and not equal as assumed here, see
\eqn{eqn:su5}. Indeed, as shown in Ref.~\cite{Cobos-Martinez:2020ynh}, the mass shift for the 
$\eta_c$ gets closer to that of the $J/\Psi$ when $SU(4)$ flavor symmetry is broken, such that
$g_{\eta_c DD^*}=(0.6/\sqrt{2})\,g_{J/\Psi DD} \simeq 0.424\,g_{J/\Psi DD}$~\cite{Cobos-Martinez:2020ynh,Lucha:2015dda}. $SU(5)$ flavor symmetry, like $SU(4)$, is also broken in nature, as 
attested by the difference is masses of the $\Upsilon$ and $\eta_b$ mesons. However, since we do 
not have an empirical value for $g_{\eta_b B B^*}$, which we can use to compute the $\eta_b$ 
self-energy, we therefore resort to $SU(5)$ symmetry and use the value for $g_{\eta_b B B^*}$ given 
in \eqn{eqn:su5};
({\bf b}) the form factors are not equal for the vertices  $\Upsilon B B$ and $\eta_b B B^*$ and we have to
readjust the cutoff values, which means $\Lambda_B\ne \Lambda_{B^*}$ and the comparisons for
the mass shifts have to be made for different values of the cutoffs. This is also a reason why we 
explore a range of values for 
$\Lambda_B$;
and ({\bf c}) at the $g_{\eta_b B B^*}^2$ order, the number of possible contractions 
to give the $BB^*$ loop in the $\eta_b$ self-energy is  
$4 \times 4 = 16$ for the isodoublet $B$ and $B^*$ fields, and this number is larger 
than that of $2 \times 2 = 4$ to give the $BB$ loop in the $\Upsilon$ self-energy
at the $g_{\Upsilon B B}^2$ order (see Eqs.~(\ref{eqn:LUBB}) and~(\ref{eqn:LetabBBast})).
This may give the larger contribution for the $\eta_b$ potential. 

The results for the mass shifts in nuclear matter, shown in Figs.~\ref{fig:upsilon_ms} 
and \ref{fig:etab_ms}, for the $\Upsilon$ and $\eta_b$, respectively, support the argument that the
nuclear medium provides attraction to these mesons and open the possibility to study the
binding of theses mesons to nuclei since the mass shifts, for both the $\Upsilon$ and $\eta_b$,
at around $\rho_B=\rho_0$, are significant. We will see in the next section that this is indeed the case
and allows for the formation of nuclear bound states for both the $\Upsilon$ and $\eta_b$, 
and furthermore we calculate the correspoding binding energies for several nuclei.

\section{\label{sec:finitenuclei} Nuclear bound states}

The results for the mass shifts of the $\Upsilon$ and $\eta_b$ in nuclear matter clearly indicate that
nuclear medium provides attraction to these mesons. Therefore, we now consider the nuclear bound
 states of the  $\Upsilon$ and $\eta_b$ mesons when these mesons have been produced nearly at 
 rest inside
  nucleus $A$ and study the following nuclei in a wide range of masses, namely 
$^{4}$He, $^{12}$C, $^{16}$O, $^{40}$Ca, $^{48}$Ca, $^{90}$Zr,  $^{197}$Au, and $^{208}$Pb. 

In the local density approximation, the bottomonium $h$ ($h=\Upsilon\,,\eta_b$) potential within nucleus
$A$ is given by 
\begin{equation}
\label{eqn:VhA}
V_{hA}(r)= \Delta m_{h}(\rho_{B}^{A}(r)),
\end{equation}
\noindent where $r$ is the distance from the center of the nucleus and $\Delta m_{h}$ is the mass
shift computed in \sect{sec:nuclmatt} for $h=\Upsilon\,,\eta_b$. 
The nuclear density distributions  $\rho_{B}^{A}(r)$ for the nuclei listed above are calculated using
the  QMC model~\cite{Saito:1996sf}, except for  $^{4}$He, which we obtain from Ref.~\cite{Saito:1997ae}.
%
%\upsilon-nucleus potentials------------------------------------------------
\begin{figure*}
  \begin{tabular}{c@{\hskip 3mm}c@{\hskip 3mm}c}
\includegraphics[scale=0.27]{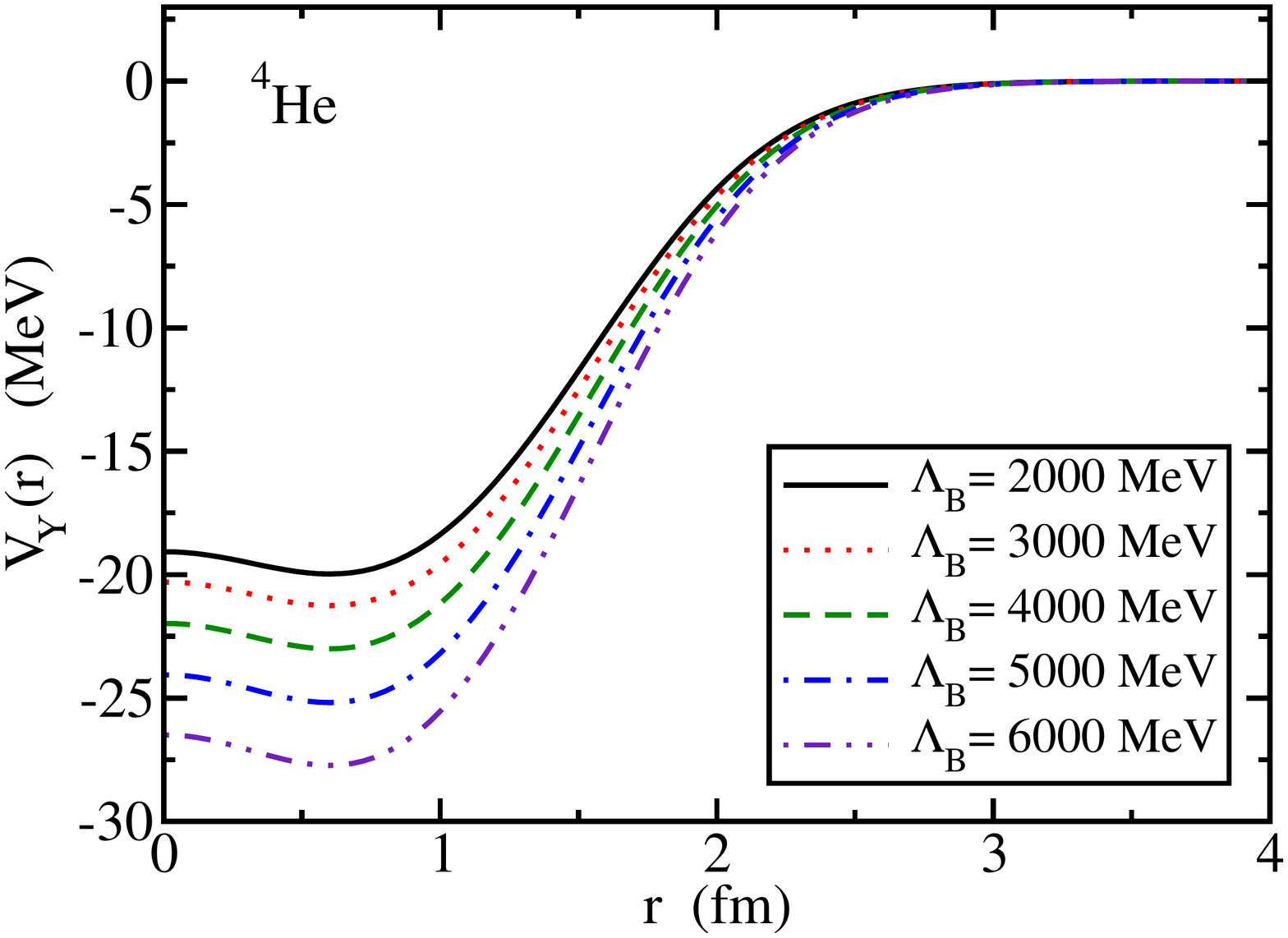} &  
\includegraphics[scale=0.27]{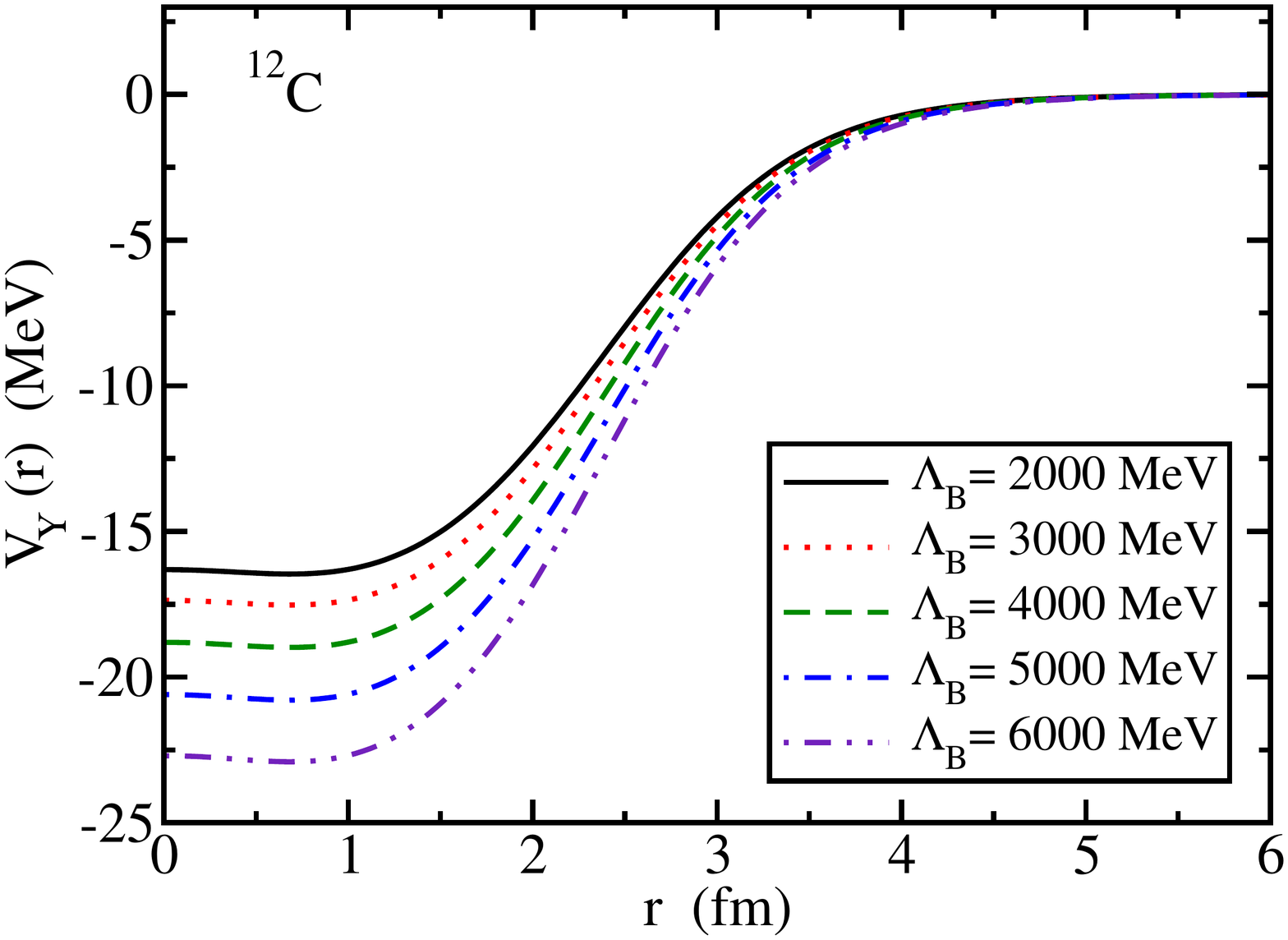} \\
\includegraphics[scale=0.27]{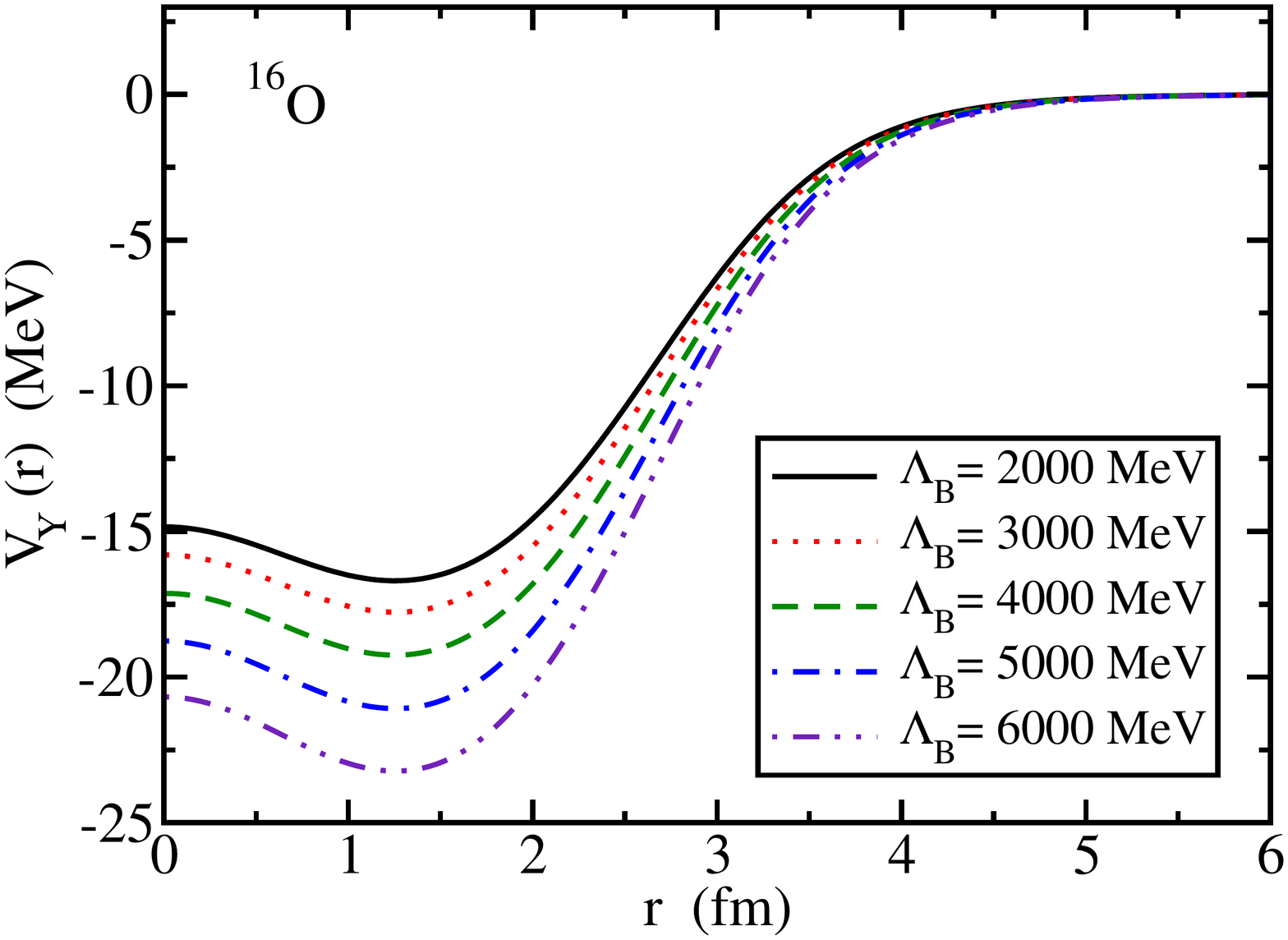} & 
\includegraphics[scale=0.27]{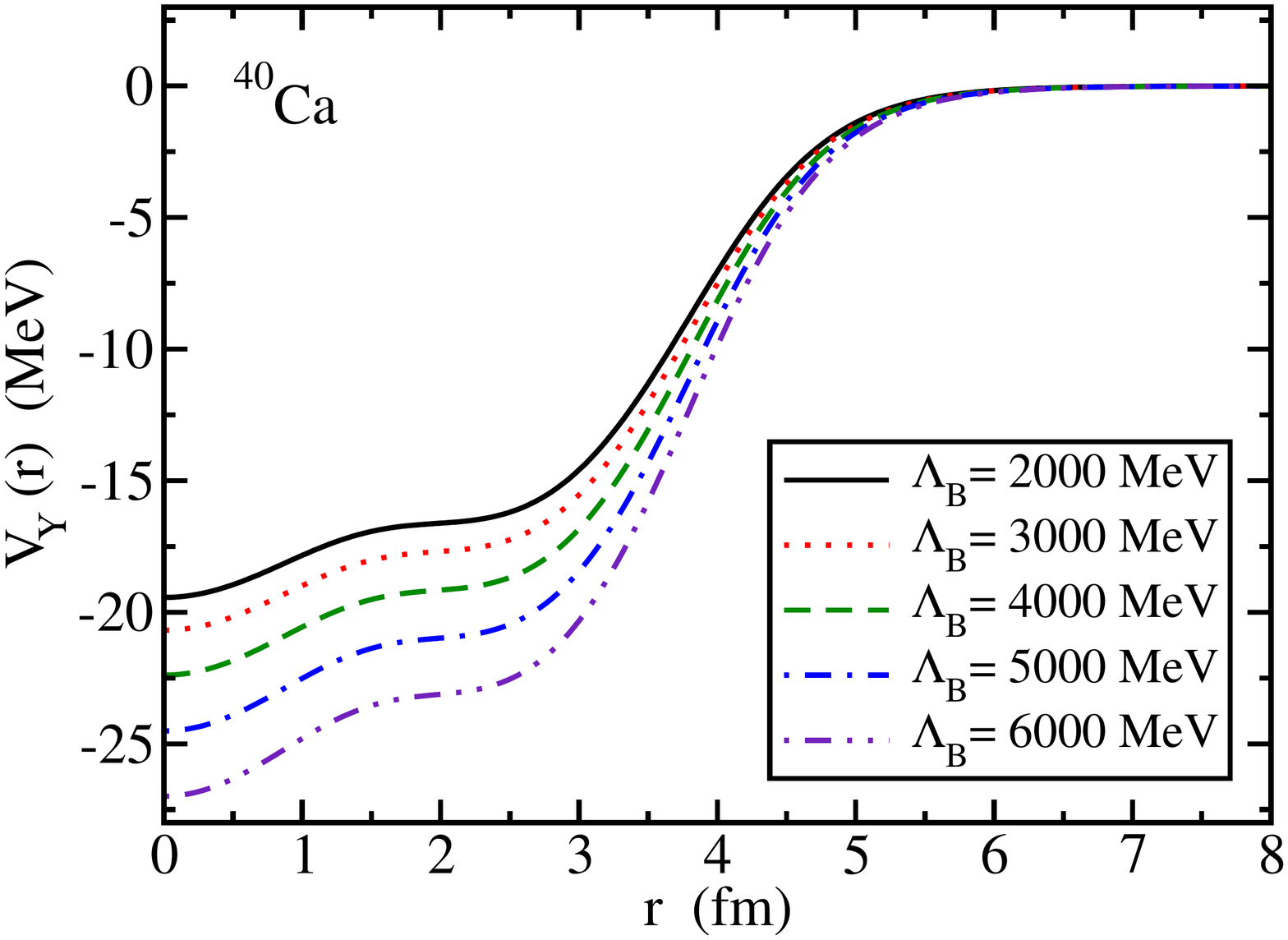} \\
\includegraphics[scale=0.27]{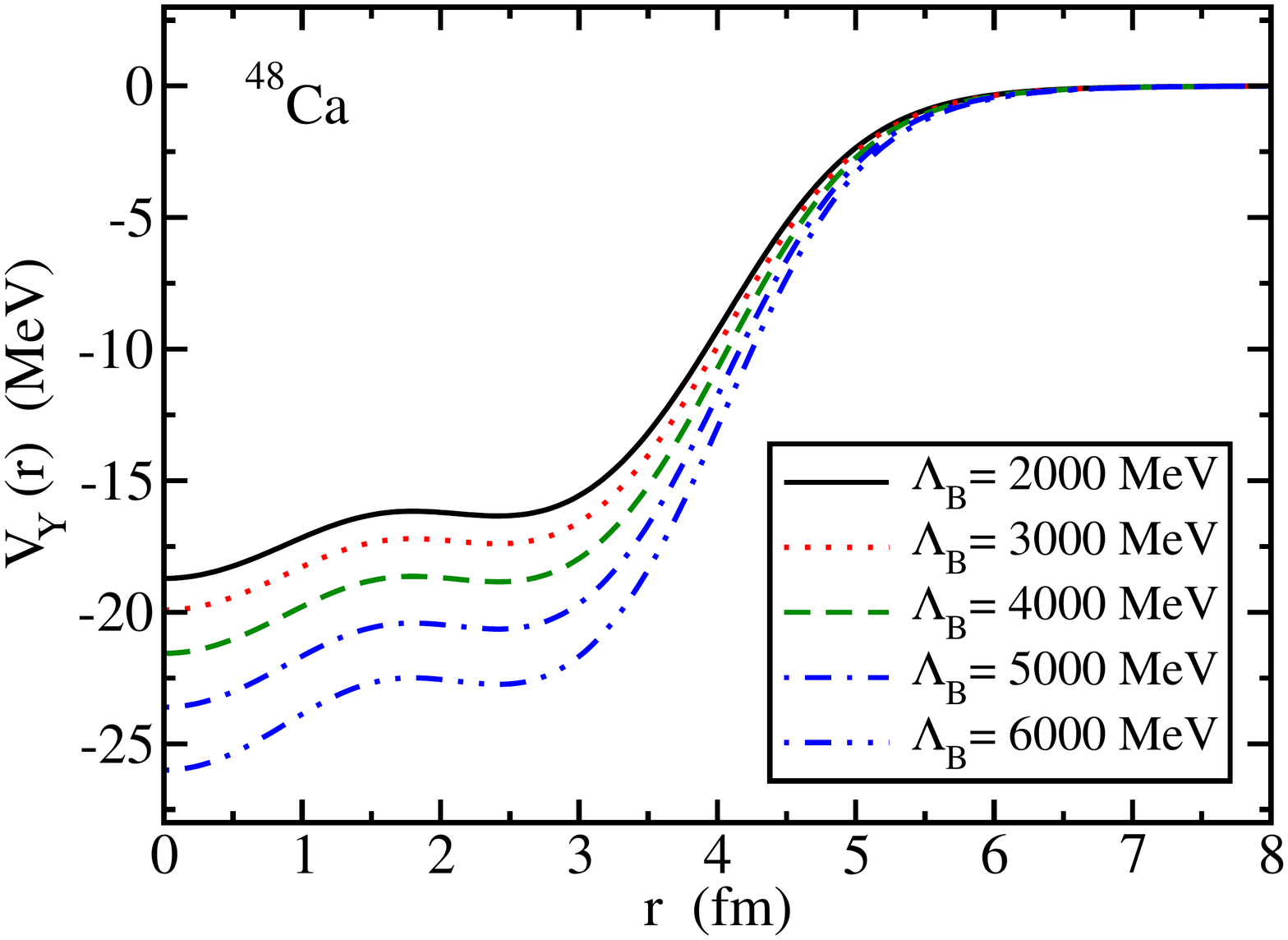} & 
\includegraphics[scale=0.27]{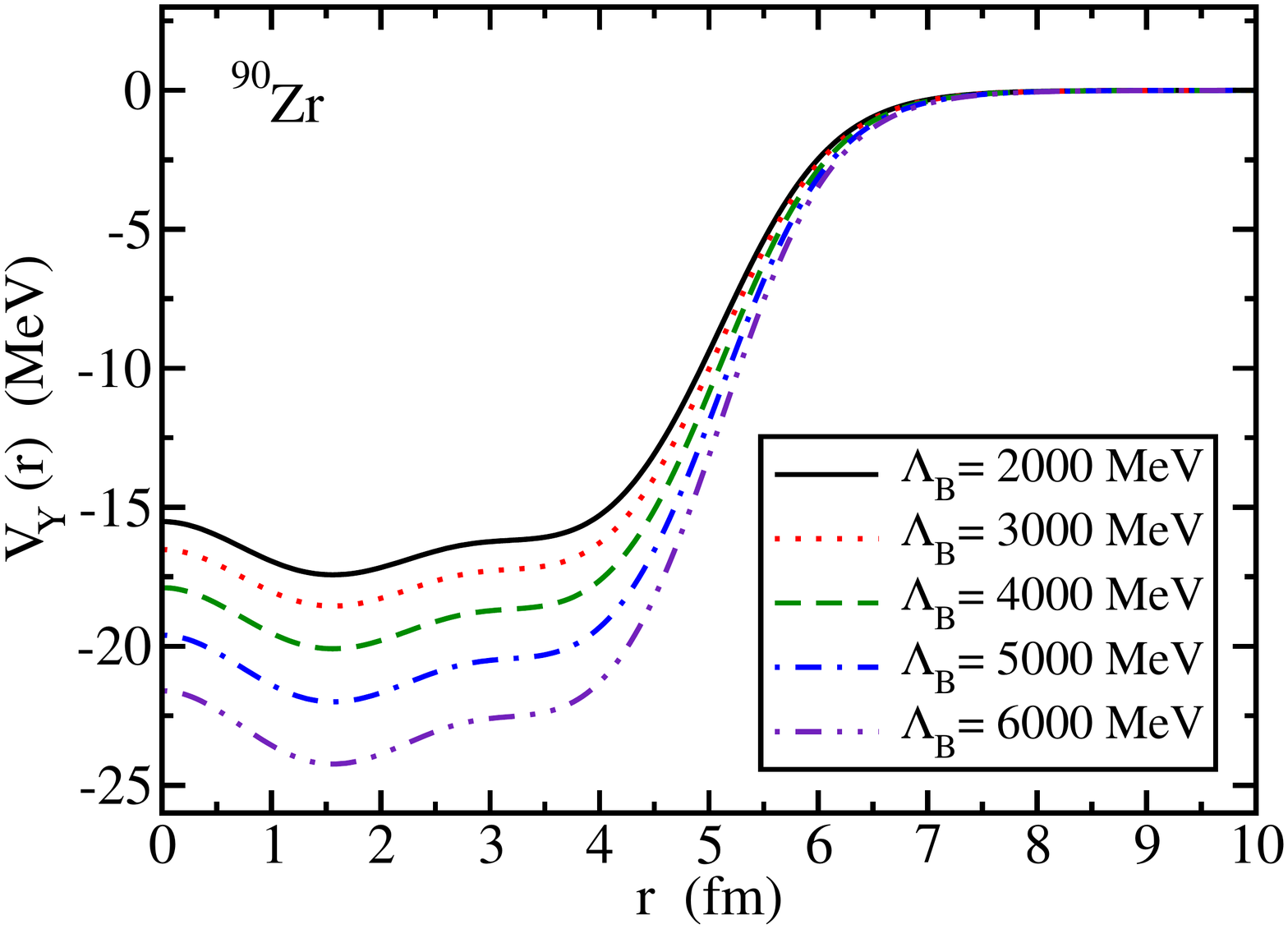} \\
\includegraphics[scale=0.27]{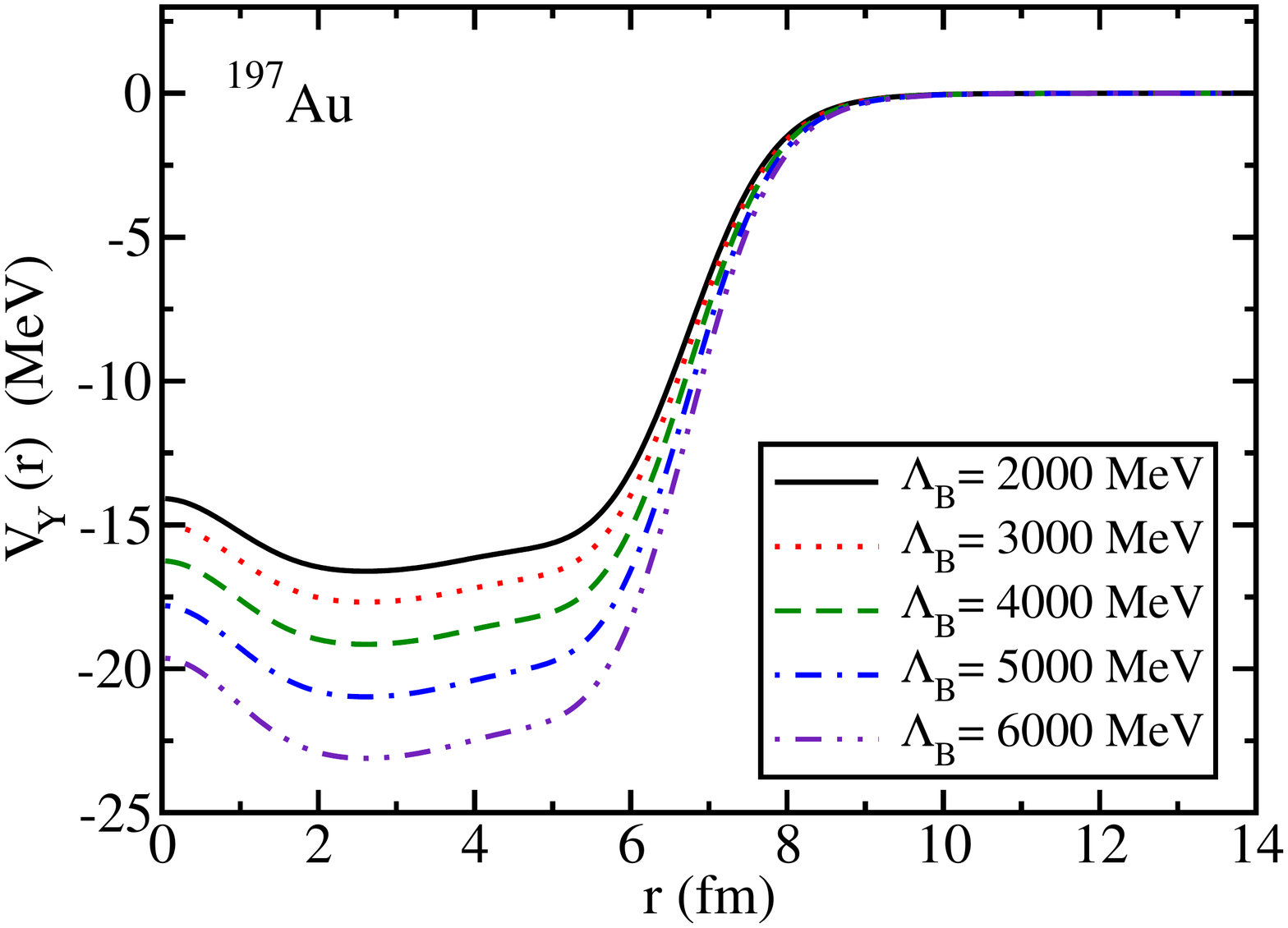} & 
\includegraphics[scale=0.27]{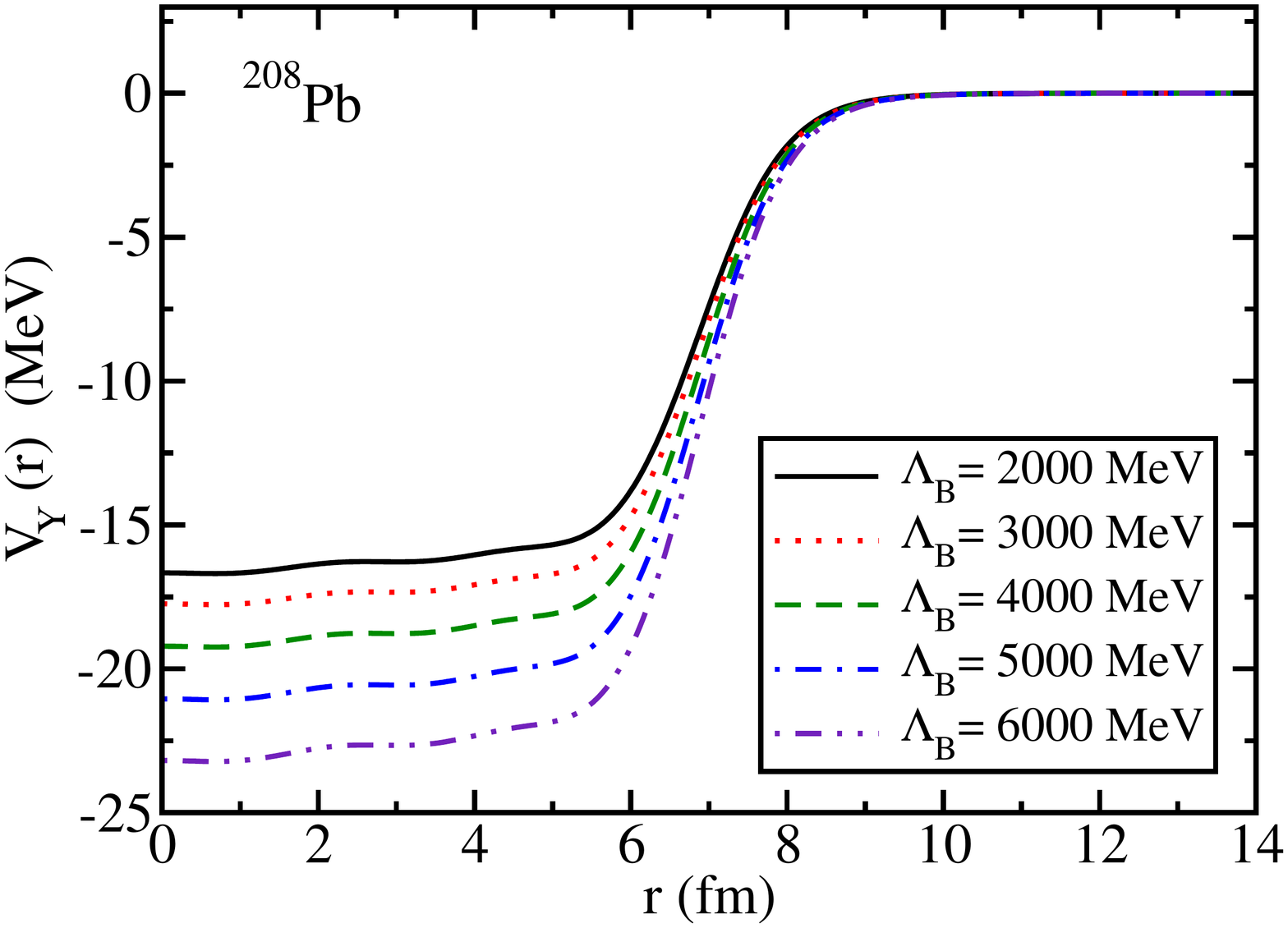} \\
  \end{tabular}
  \caption{\label{fig:VUpsilonA}
    $\Upsilon$-nucleus  potentials for various nuclei and values of the
    cutoff parameter $\Lambda_{B}$.}
\end{figure*}
%--------------------------------------------------------------------
%
%\upsilon-nucleus potentials------------------------------------------------
\begin{figure*}
  \begin{tabular}{c@{\hskip 3mm}c@{\hskip 3mm}c}
\includegraphics[scale=0.27]{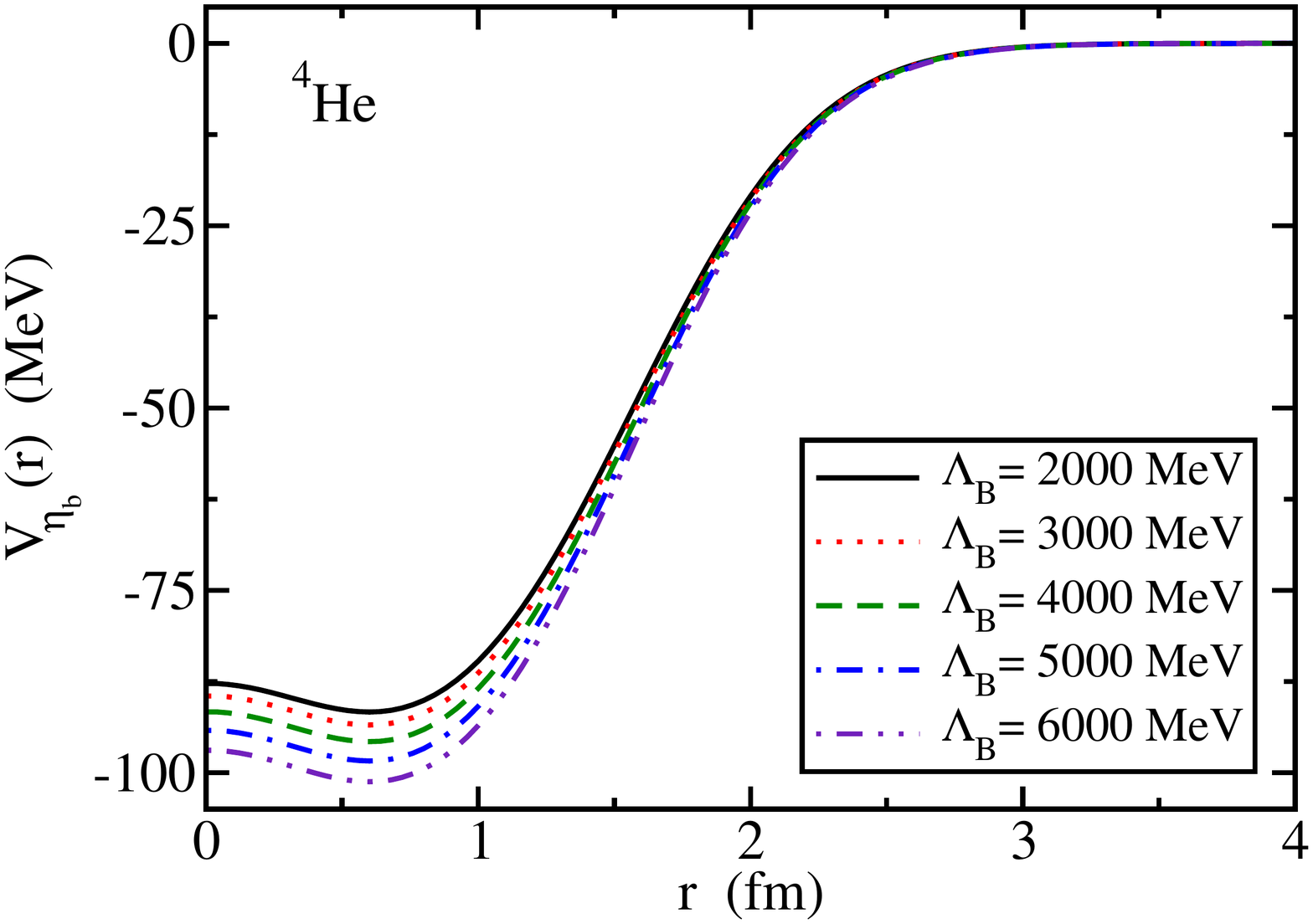} &  
\includegraphics[scale=0.27]{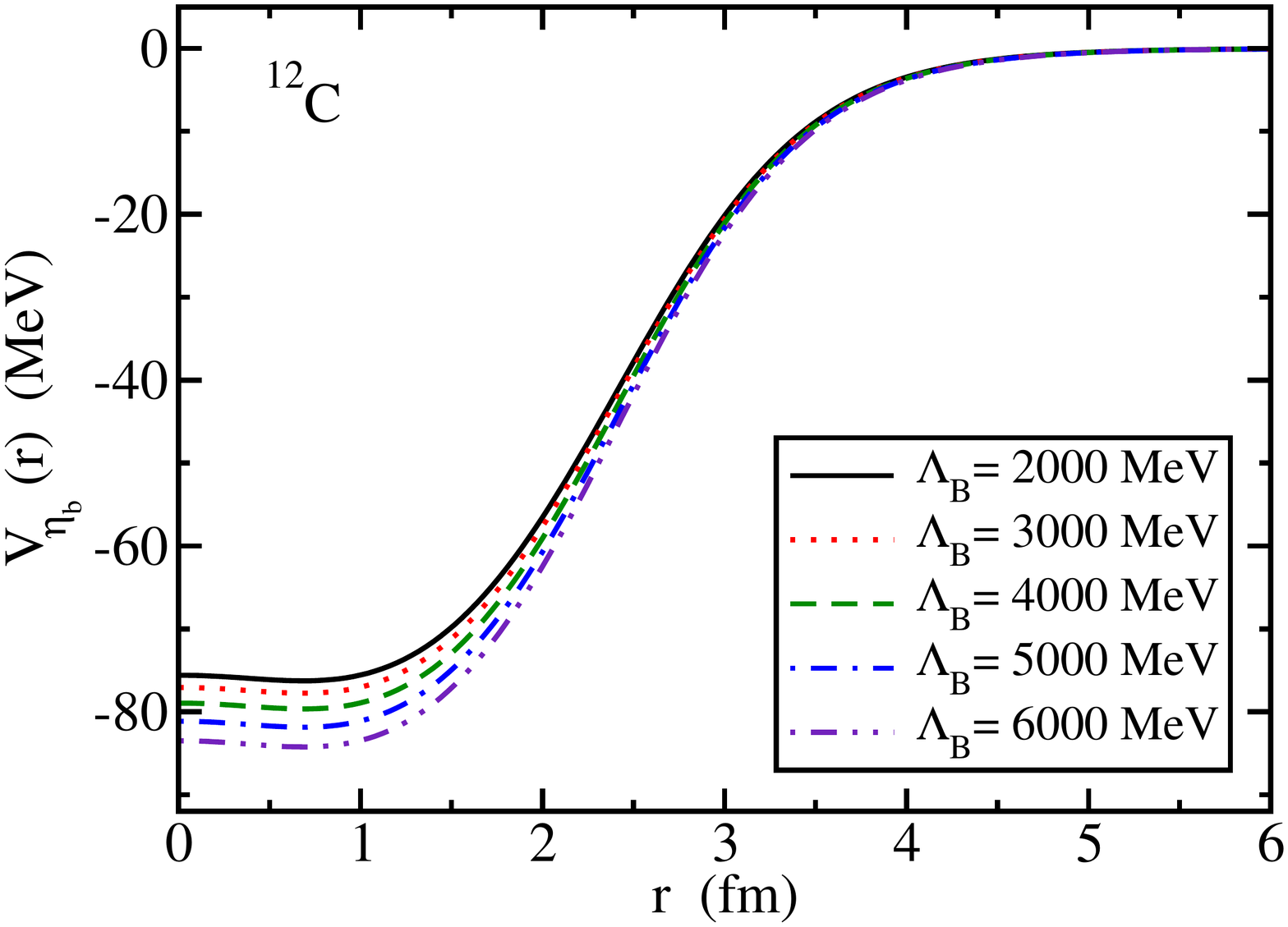} \\
\includegraphics[scale=0.27]{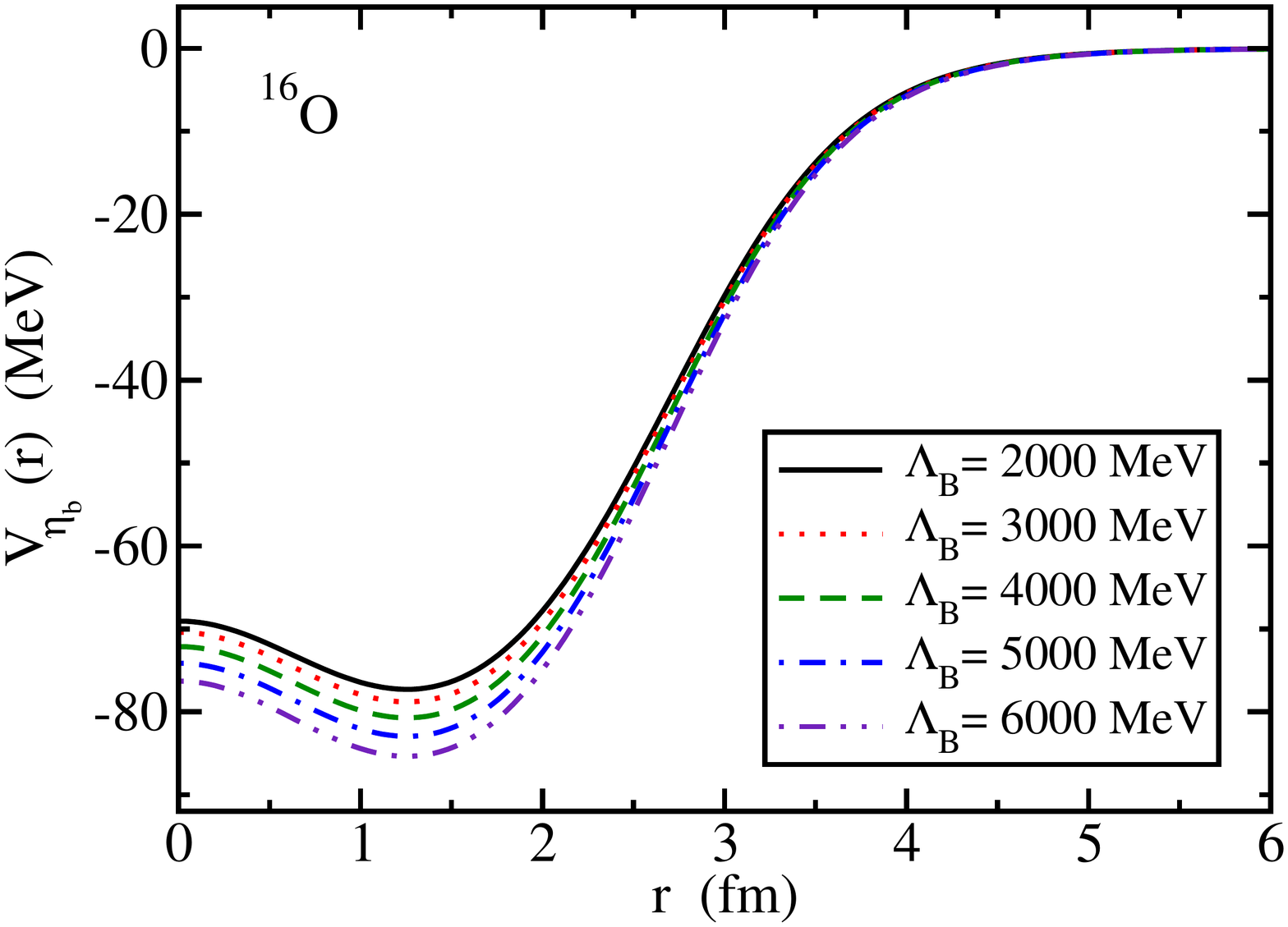} & 
\includegraphics[scale=0.27]{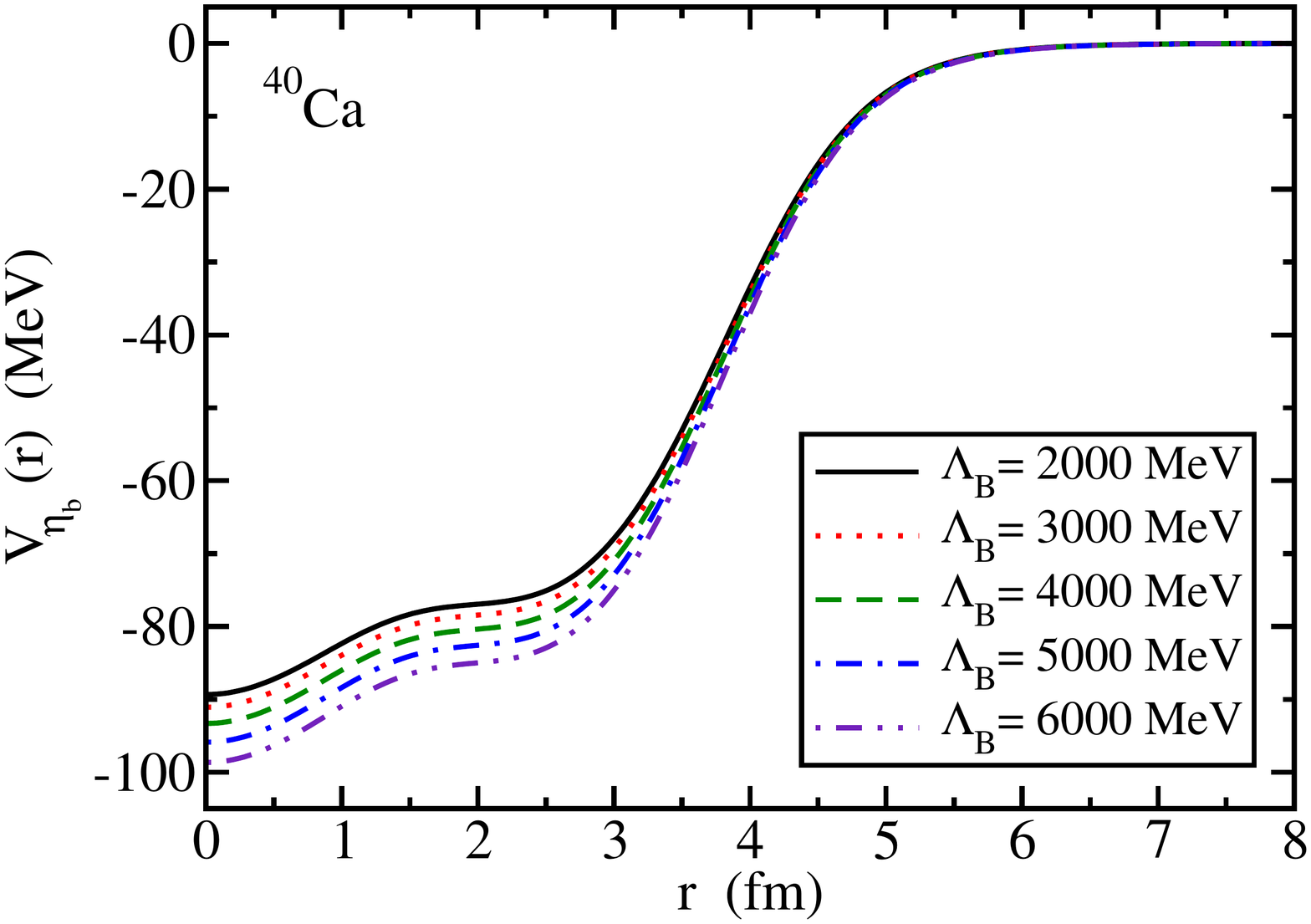} \\
\includegraphics[scale=0.27]{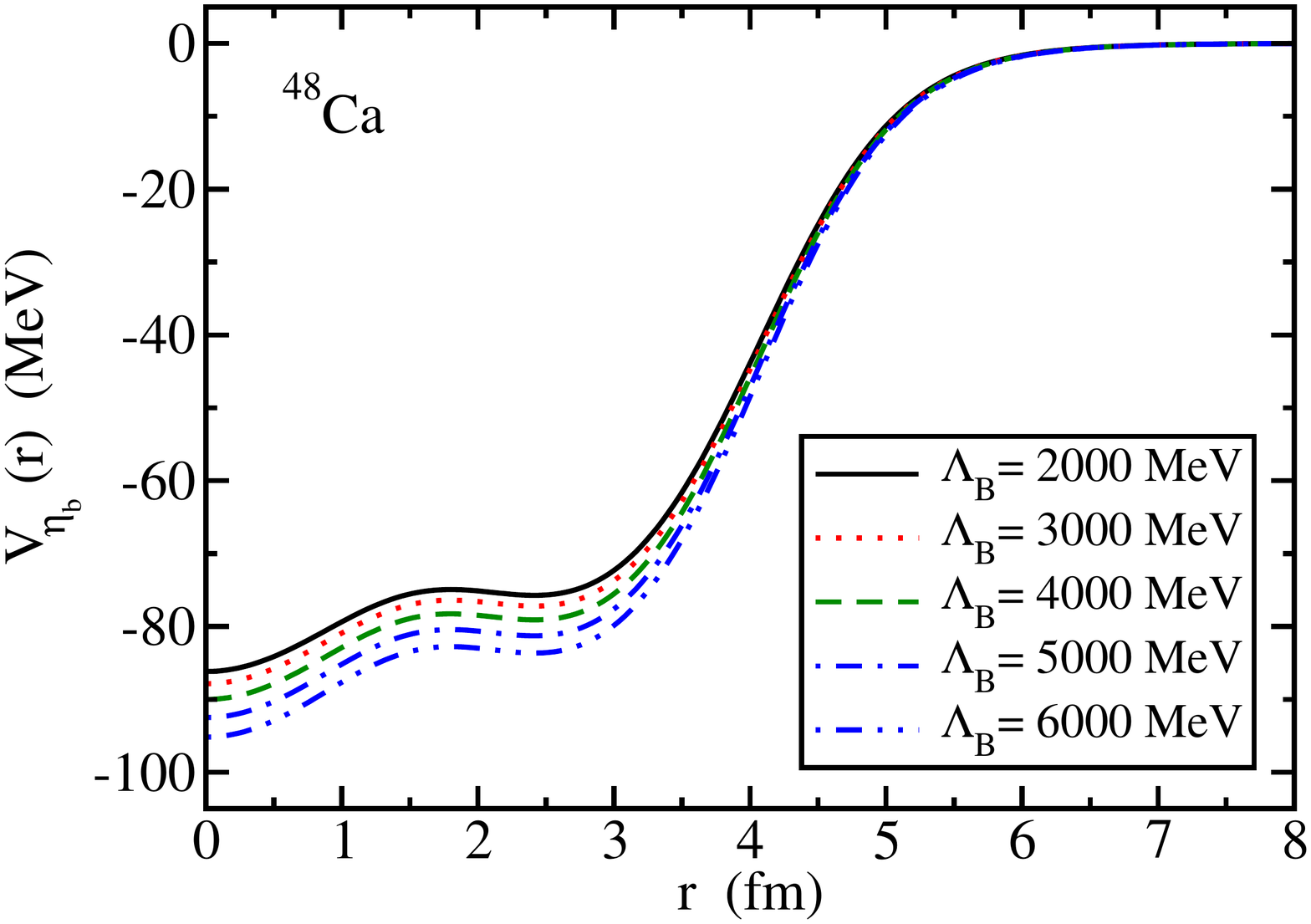} & 
\includegraphics[scale=0.27]{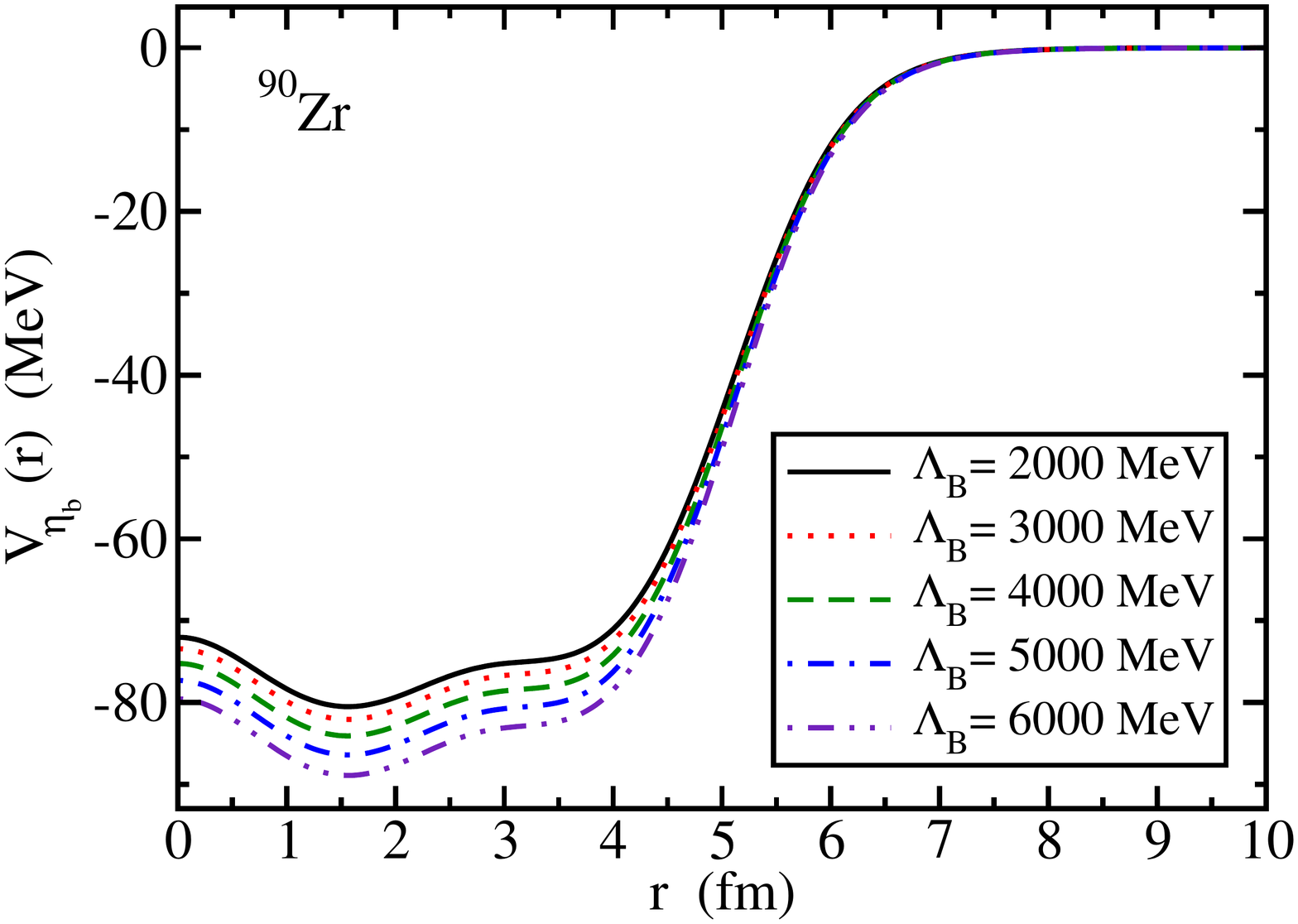} \\
\includegraphics[scale=0.27]{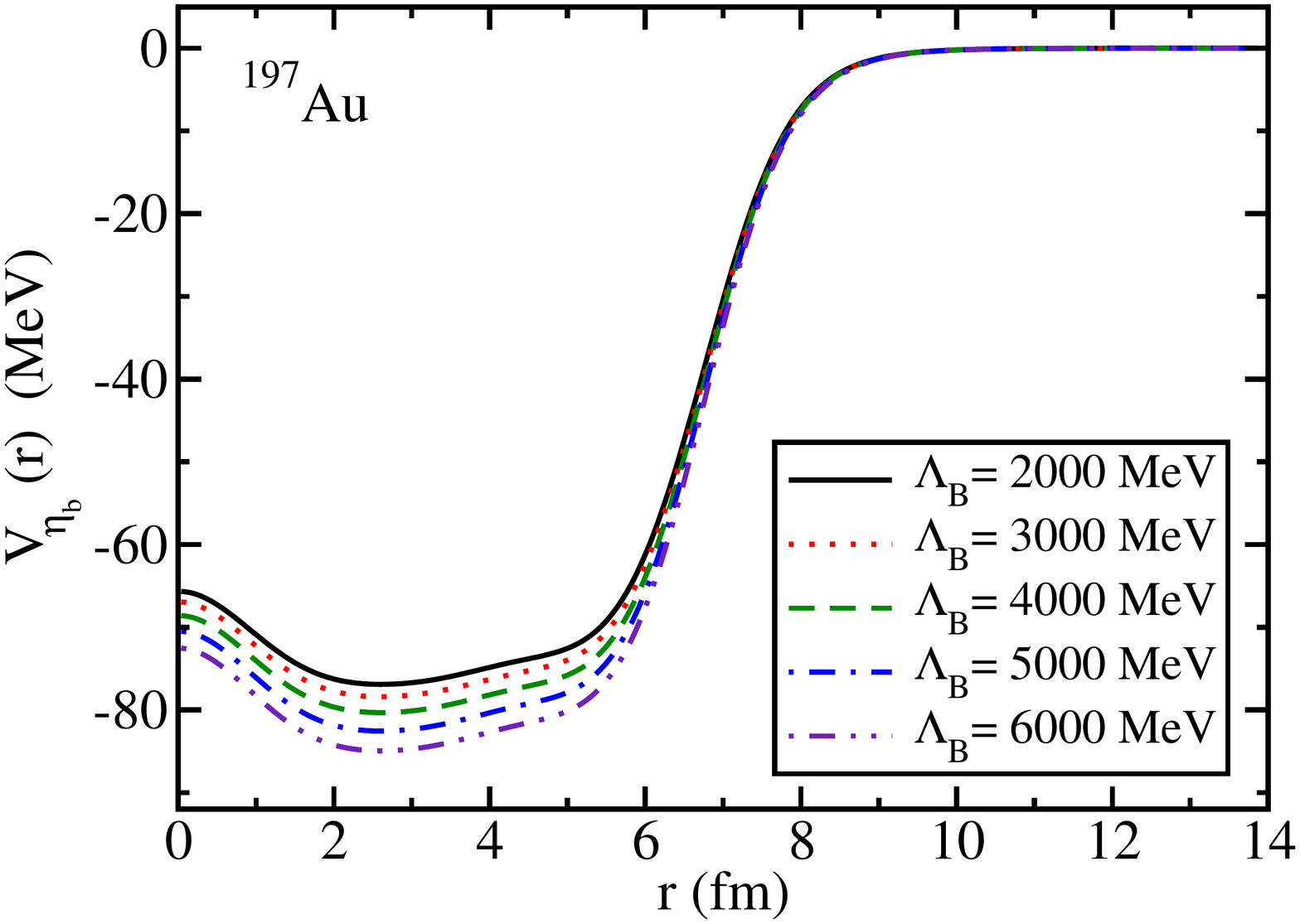} & 
\includegraphics[scale=0.27]{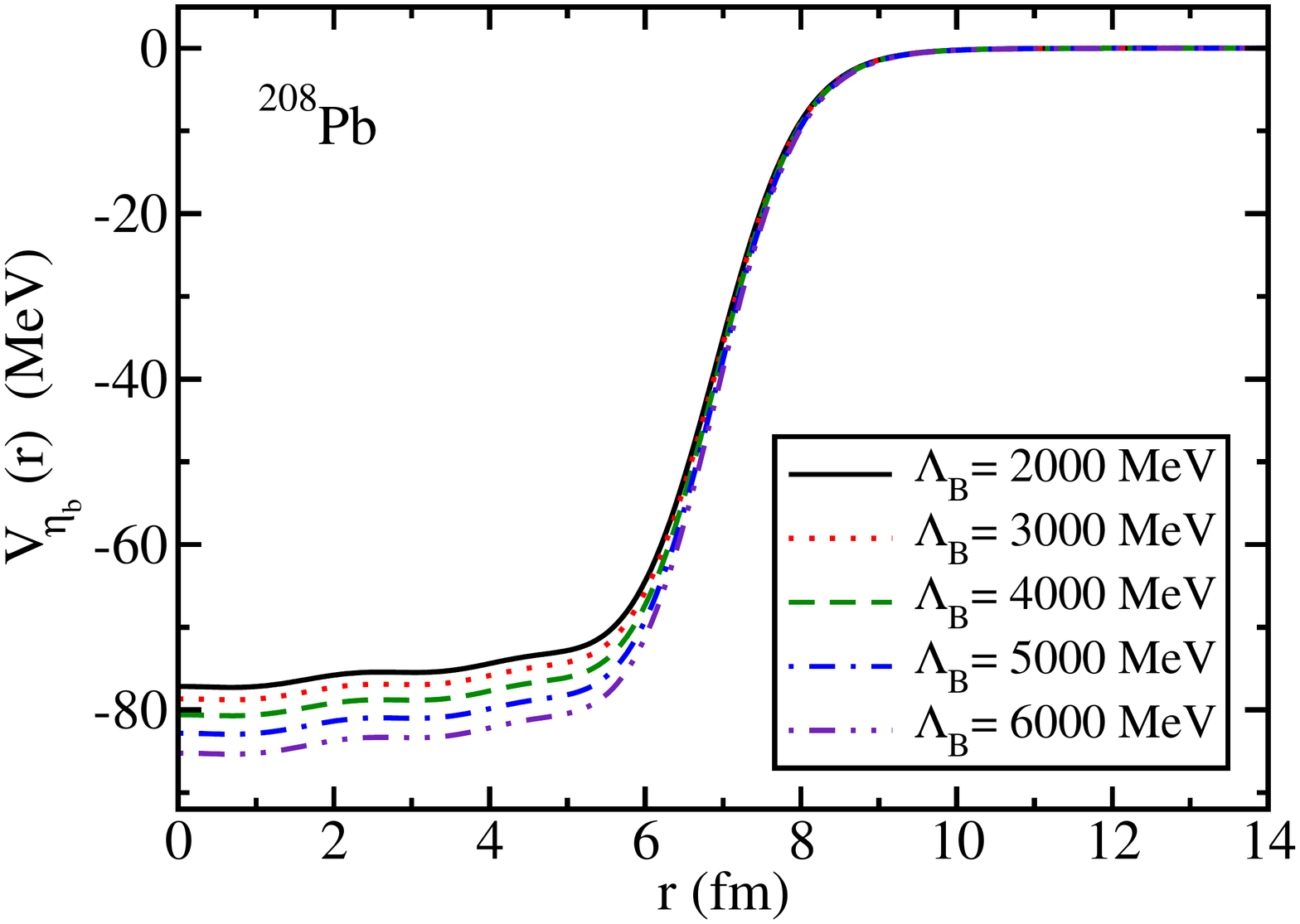} \\
  \end{tabular}
  \caption{\label{fig:VetabA}
    $\eta_b$-nucleus  potentials for various nuclei and values of the
    cutoff parameter $\Lambda_{B}$.}
\end{figure*}
%--------------------------------------------------------------------
%
In Figs.~\ref{fig:VUpsilonA} and \ref{fig:VetabA} we present the bottomonium $h$-nucleus potentials
for the eigth nuclei mentioned above and the sames values of the cutoff parameter $\Lambda_{B}$
that were used in the computation of the mass shifts in \sect{sec:nuclmatt}. 
We can see from Figs.~\ref{fig:VUpsilonA} and \ref{fig:VetabA}  that the $V_{hA}$ potentials, 
for $h=\Upsilon$ and $\eta_b$,  respectively, are attractive for all nuclei and all values of the cutoff mass
parameter.
However, for each nuclei, the depth of the potential depends on the value of the  cutoff parameter, 
being more attractive the larger $\Lambda_{B}$ is. This dependence is expected and is, indeed, an
uncertainty in the results obtained in our approach. 

We now compute the bottomonium $h$-nucleus bound state energies for the potentials shown in
Figs.~\ref{fig:VUpsilonA} and \ref{fig:VetabA} by solving the Klein-Gordon equation for these 
potentials. In order to apply the Klein-Gordon equation to obtain the $\Upsilon$-nucleus single particle
 energies, since the $\Upsilon$ is a spin-1 particle, we make an approximation where the transverse and
 longitudinal components in the Proca equation are expected to be very similar for an $\Upsilon$ at
 rest, hence it is reduced to one component, which corresponds to the Klein-Gordon equation.
 
We treat the bottomonium $h$-nucleus potential as a scalar and add it to the mass term in the 
Klein-Gordon  equation
\begin{equation}
\label{eqn:kg}
\left(-\nabla^{2} + \left(m +V_{hA}(\vec{r})\right)^2\right)\phi_{h}(\vec{r})
= \mathcal{E}^{2}\phi_{h}(\vec{r}),
\end{equation}
\noindent where $h=\Upsilon\,,\eta_b$, $m=m_{h}m_{A}/(m_{h}+m_{A})$ is the reduced mass of 
the bottomonium  $h$-nucleus system with $m_{h}$ $(m_{A}$) the mass of bottomonium $h$ 
(nucleus $A$) in  vacuum, and $V_{hA}(\vec{r})$ is the bottomonium $h$-nucleus potential given
in  \eqn{eqn:VhA} and shown in Figs.~\ref{fig:VUpsilonA} and \ref{fig:VetabA}. 

We note that in previous works we have approximated \eqn{eqn:kg} by ignoring the $V_{hA}^2$ term.
We now solve the full Klein-Gordon equation \eqn{eqn:kg} using momentum space methods. Here,
the Klein-Gordon equation is first converted to a momentum space representation via a Fourier transform,
followed by a partial wave decomposition.  For a given value of angular momentum $l$, the 
eigenvalues $\mathcal{E}_{nl}$ of the resulting equation are found by the inverse iteration eigenvalue algorithm. 

The bound state energies ($E$) of the bottomonium $h$-nucleus system, given by 
$E_{nl}=\mathcal{E}_{nl}-m$, are listed in Tables~\ref{tab:upsilon-A-BSE} and \ref{tab:etab-A-BSE}. 
In \tab{tab:upsilon-A-BSE} we show the $\Upsilon$-nucleus bound state energies for all
nuclei listed at the beginning of this section and the same  range of values for the cutoff mass 
parameter as used in the mass shift calculation.
For each nucleus we have listed only a few bound states, since the number of bound states increases 
with the mass of the nucleus and for the heaviest of these the number of bound states quiet large. 
For example, for the heaviest nucleus we have $\sim$ 70 states.
%
%Bound state energies from the Klein-Gordon equation--------------------------
\begin{table}[h]
  \caption{\label{tab:upsilon-A-BSE} $^{4}_{\Upsilon}\text{A}$ bound state
  energies for several nuclei $A$. All dimensioned quantities are in MeV.}
\begin{center}
\scalebox{0.8}{
\begin{tabular}{ll|r|r|r|r|r}
  \hline \hline
  & & \multicolumn{5}{c}{Bound state energies} \\
  \hline
& $n\ell$ & $\Lambda_{B}=2000$ & $\Lambda_{B}=3000$ & $\Lambda_{B}= 4000$ &
$\Lambda_{B}= 5000$ & $\Lambda_{B}= 6000$ \\
\hline
$^{4}_{\Upsilon}\text{He}$
& 1s &  -5.6 &  -6.4 & -7.5 & -9.0 & -10.8 \\
\hline
$^{12}_{\Upsilon}\text{C}$
& 1s & -10.6 & -11.6 & -12.8 & -14.4 & -16.3 \\
& 1p & -6.1 & -6.8 & -7.9 & -9.3 & -10.9 \\
& 1d & -1.5 & -2.1 & -2.9 & -4.0 & -5.4 \\
& 2s & -1.6 & -2.1 & -2.8 & -3.8 & -5.1 \\
\hline
$^{16}_{\Upsilon}\text{O}$
& 1s & -11.9 & -12.9 & -14.2 & -15.8 & -17.8 \\
& 1p & -8.3 & -9.2 & -10.4 & -11.9 & -13.7 \\
& 1d & -4.4 & -5.1 & -6.2 & -7.5 & -9.2 \\
& 2s & -3.7 & -4.4 & -5.4 & -6.7 & -8.3 \\
& 1f & n & -0.9 & -1.8 & -2.9 & -4.3 \\
\hline
$^{40}_{\Upsilon}\text{Ca}$
& 1s & -15.5 & -16.6 & -18.2 & -20.0 & -22.3 \\
& 1p & -13.3 & -14.4 & -15.9 & -17.7 & -19.8 \\
& 1d & -10.8 & -11.9 & -13.3 & -15.0 & -17.1 \\
& 2s & -10.3 & -11.3 & -12.7 & -14.4 & -16.4 \\
& 1f & -8.1 & -9.1 & -10.4 & -12.1 & -14.0 \\
\hline
$^{48}_{\Upsilon}\text{Ca}$
& 1s & -15.3 & -16.4 & -17.9 & -19.7 & -21.8 \\
& 1p & -13.5 & -14.6 & -16.0 & -17.8 & -19.9 \\
& 1d & -11.4 & -12.4 & -13.8 & -15.6 & -17.6 \\
& 2s & -10.8 & -11.8 & -13.2 & -14.9 & -16.9 \\
& 1f & -9.1 & -10.1 & -11.4 & -13.1 & -15.0 \\
\hline
$^{90}_{\Upsilon}\text{Zr}$
& 1s & -15.5 & -16.6 & -18.1 & -19.9 & -22.0 \\
& 1p & -14.5 & -15.5 & -17.0 & -18.8 & -20.9 \\
& 1d & -13.2 & -14.2 & -15.7 & -17.4 & -19.5 \\
& 2s & -12.7 & -13.8 & -15.2 & -16.9 & -19.0 \\
& 1f & -11.7 & -12.7 & -14.1 & -15.9 & -17.9 \\
\hline
$^{197}_{\Upsilon}\text{Au}$
& 1s & -15.3 & -16.3 & -17.7 & -19.4 & -21.5 \\
& 1p & -14.7 & -15.8 & -17.2 & -18.9 & -20.9 \\
& 1d & -14.0 & -15.0 & -16.4 & -18.1 & -20.1 \\
& 2s & -13.7 & -14.7 & -16.0 & -17.8 & -19.8 \\
& 1f & -13.2 & -14.2 & -15.6 & -17.3 & -19.3 \\
\hline
$^{208}_{\Upsilon}\text{Pb}$
& 1s & -15.7 & -16.8 & -18.2 & -20.0 & -22.1 \\
& 1p & -15.2 & -16.2 & -17.7 & -19.4 & -21.5 \\
& 1d & -14.5 & -15.5 & -16.9 & -18.7 & -20.8 \\
& 2s & -14.1 & -15.2 & -16.6 & -18.3 & -20.4 \\
& 1f & -13.6 & -14.7 & -16.1 & -17.8 & -19.9 \\
\hline
\end{tabular}
}
\end{center}
\end{table}
%-----------------------------------------------------------------------------
%
In \tab{tab:etab-A-BSE} we show the $\eta_b$-nucleus bound state energies for the same
nuclei and range of values of the cutoff mass parameter as in \tab{tab:upsilon-A-BSE}.
Furthermore, as in the case of the case of the  $\Upsilon$-nucleus bound state energies, for
each nucleus we have listed only a few bound states. For the heaviest nucleus we have $\sim$ 
200 states and clearly is not practical to show them all.

%
%Bound state energies from the Klein-Gordon equation--------------------------
\begin{table}[h]
  \caption{\label{tab:etab-A-BSE} $^{4}_{\eta_b}\text{A}$ bound state
  energies for several nuclei $A$. All dimensioned quantities are in MeV.}
\begin{center}
\scalebox{0.8}{
\begin{tabular}{ll|r|r|r|r|r}
  \hline \hline
  & & \multicolumn{5}{c}{Bound state energies} \\
  \hline
& $n\ell$ & $\Lambda_{B}=2000$ & $\Lambda_{B}=3000$ & $\Lambda_{B}= 4000$ &
$\Lambda_{B}= 5000$ & $\Lambda_{B}= 6000$ \\
\hline
$^{4}_{\eta_b}\text{He}$
& 1s & -63.1 & -64.7 & -66.7 & -69.0 & -71.5 \\ 
& 1p & -40.6 & -42.0 & -43.7 & -45.8 & -48.0 \\ 
& 1d & -17.2 & -18.3 & -19.7 & -21.4 & -23.2 \\ 
& 2s & -15.6 & -16.6 & -17.9 & -19.4 & -21.1 \\ 
\hline
$^{12}_{\eta_b}\text{C}$
& 1s & -65.8 & -67.2 & -69.0 & -71.1 & -73.4 \\
& 1p & -57.0 & -58.4 & -60.1 & -62.1 & -64.3 \\
& 1d & -47.5 & -48.8 & -50.4 & -52.3 & -54.4 \\
& 2s & -46.3 & -47.5 & -49.1 & -51.0 & -53.0 \\
& 1f & -37.5 & -38.7 & -40.2 & -42.0 & -43.9 \\
\hline
$^{16}_{\eta_b}\text{O}$
& 1s & -67.8 & -69.2 & -71.0 & -73.1 & -75.4 \\
& 1p & -61.8 & -63.2 & -64.9 & -67.0 & -69.2 \\
& 1d & -54.9 & -56.2 & -57.9 & -59.9 & -62.0 \\
& 2s & -53.2 & -54.6 & -56.3 & -58.2 & -60.3 \\
& 1f  & -47.3 & -48.6 & -50.2 & -52.1 & -54.2 \\
\hline
$^{40}_{\eta_b}\text{Ca}$
& 1s & -79.0 & -80.6 & -82.6 & -85.0 & -87.5 \\
& 1p & -75.4 & -77.0 & -79.0 & -81.4 & -83.9 \\
& 1d & -71.4 & -73.0 & -74.9 & -77.2 & -79.7 \\
& 2s & -70.5 & -72.0 & -74.0 & -76.3 & -78.8 \\
& 1f & -67.0 & -68.5 & -70.4 & -72.7 & -75.1 \\
\hline
$^{48}_{\eta_b}\text{Ca}$
& 1s & -76.7 & -78.2 & -80.2 & -82.5 & -85.0 \\
& 1p & -74.0 & -75.5 & -77.4 & -79.7 & -82.1 \\
& 1d & -70.8 & -72.3 & -74.2 & -76.4 & -78.8 \\
& 2s & -69.9 & -71.4 & -73.3 & -75.5 & -77.9 \\
& 1f & -67.2 & -68.6 & -70.6 & -72.8 & -75.1 \\
\hline
$^{90}_{\eta_b}\text{Zr}$
& 1s & -75.5 & -77.0 & -78.9 & -81.1 & -83.5 \\
& 1p & -74.1 & -75.6 & -77.5 & -79.7 & -82.1 \\ 
& 1d & -72.3 & -73.8 & -75.7 & -77.9 & -80.2 \\ 
& 2s & -71.6 & -73.0 & -74.9 & -77.1 & -79.5 \\ 
& 1f & -70.2 & -71.7 & -73.6 & -75.8 & -78.1 \\ 
\hline
$^{197}_{\eta_b}\text{Au}$
& 1s & -72.8 & -74.2 & -76.1 & -78.2 & -80.5 \\ 
& 1p & -72.3 & -73.7 & -75.6 & -77.7 & -80.0 \\ 
& 1d & -71.3 & -72.8 & -74.6 & -76.7 & -79.0 \\ 
& 2s & -70.7 & -72.1 & -74.0 & -76.1 & -78.4 \\ 
& 1f & -70.2 & -71.7 & -73.5 & -75.6 & -77.9 \\ 
\hline
$^{208}_{\eta_b}\text{Pb}$
& 1s & -74.7 & -76.2 & -78.1 & -80.3 & -82.6 \\ 
& 1p & -74.2 & -75.7 & -77.5 & -79.7 & -82.1 \\ 
& 1d & -73.2 & -74.7 & -76.6 & -78.8 & -81.1 \\ 
& 2s & -72.7 & -74.1 & -76.0 & -78.2 & -80.5 \\ 
& 1f & -72.1 & -73.6 & -75.5 & -77.6 & -80.0 \\ 
\hline
\end{tabular}
}
\end{center}
\end{table}
%-----------------------------------------------------------------------------
%
We can now give some general conclusion concerning the results given in 
Tables~\ref{tab:upsilon-A-BSE}  and \ref{tab:etab-A-BSE}. These results show that the 
$\Upsilon$ and $\eta_b$ mesons are expected to form bound states with all the nuclei studied, independent of the value  of the cutoff parameter $\Lambda_{B}$. However, the particular values 
for the bound state energies are dependent on the cutoff parameter, increasing in absolute value as the cutoff parameter increases.
This dependence was  expected from the behavior of the bottomonium $h$-nucleus potentials, 
since these are more attractive for larger values of the cutoff parameter.
Note also that bottomonium $h$ binds more strongly to heavier nuclei and therefore a richer
spectrum is expected for these nuclei. \\

\subsection{Discussion of  the $\Upsilon$ and $\eta_b$ sigle particle energies}

The discussion of the mass shifts results for the $\Upsilon$ and $\eta_b$ carried out in section
\ref{sec: mass shift discussion} can be translated to the $\Upsilon$ and $\eta_b$ sigle particle energies.
From Tables ~\ref{tab:upsilon-A-BSE}  and \ref{tab:etab-A-BSE}, we see that the 
bound state energies for the $\eta_b$ are larger than those of the $\Upsilon$ for the same
nuclei and range of cutoff values explored. As before, these differences are 
probably due to two reasons: ({\bf a}) the couplings $g_{\eta_b BB^*}$ and $g_{\Upsilon BB}$ 
are very different. Indeed, the results obtained in Ref.~\cite{Cobos-Martinez:2020ynh} on the
$\eta_c$ nuclear bound state energies are closer to those the $J/\Psi$ when the $SU(4)$ flavor symmetry is broken, such that $g_{\eta_c DD^*}=(0.6/\sqrt{2})\,g_{J/\Psi DD} \simeq 0.424\,g_{J/\Psi DD}$~\cite{Cobos-Martinez:2020ynh,Lucha:2015dda}. 
Thus a reduced coupling $g_{\eta_b BB^*}$ can bring the $\eta_b$ nuclear bound state energies closer to those the $\Upsilon$, since the $\eta_b$ self-energy is proportional to $g_{\eta_b BB^*}^2$.
({\bf b}) the form factors are not equal for the vertices  $\Upsilon B B$ and $\eta_b B B^*$ and we have to
readjust the cutoff values, which means $\Lambda_B\ne \Lambda_{B^*}$, and the comparisons for
the mass shifts have to be made for different values for the cutoff parameters.

\section{\label{sec:summary} Summary and discussion}

We have calculated the $\Upsilon$- and $\eta_b$-nucleus bound state energies for
various nuclei neglecting any possible effects of the widths
and various values of the cutoff parameter $\Lambda_B$ that was introduced 
to regularize the divergent integral in the self energies for these mesons.
The bottomonium $h$-nucleus potentials were calculated using a local density approximation, 
with the inclusion of the $BB$ ($BB^*$) meson loop in the $\Upsilon$ ($\eta_c$) self-energy.
The nuclear density distributions and the in-medium $B$ and 
$B^{*}$ meson masses were calculated using  the quark-meson coupling model. 
Using the bottomonium $h$ potentials in nuclei, we have solved the Klein-Gordon 
equation and obtained bottomonium $h$-nucleus bound state energies.

Our results show that the  $\Upsilon$ and $\eta_b$ mesons are expected to form bound states with all the nuclei studied, independent of the value  of the cutoff parameter $\Lambda_{B}$. However, the particular values  for the bound state energies are dependent on cutoff parameter $\Lambda_{B}$.
The sensitivity of our results to the  cutoff parameter $\Lambda_{B}$ has also been explored. 
However, an study needs to be done where we use a more properly determined coupling 
$g_{\eta_b BB^*}$  and different functional forms for the form factors. 
Furthermore, it is certainly necessary to include the possible effects of the 
widths for the $\Upsilon$ and $\eta_b$.
Such elaborated studies are underway and will be reported elsewhere.

\section*{Acknowledgements}
\noindent
We thank Prof.~Tomoi Koide for useful conversations.
GNZ was supported in part by the 
Coordena\c{c}\~ao de Aperfeiçoamento de Pessoal de N\'ivel Superior 
- Brazil (CAPES), and KT was supported by the Conselho Nacional de Desenvolvimento
Cient\'{i}fico e Tecnol\'{o}gico (CNPq)
Process, No.~313063/2018-4, and No.~426150/2018-0,
and Funda\c{c}\~{a}o de Amparo \`{a} Pesquisa do Estado
de S\~{a}o Paulo (FAPESP) Process, No.~2019/00763-0,
and this work was also part of the projects, Instituto Nacional de Ci\^{e}ncia e
Tecnologia --- Nuclear Physics and Applications (INCT-FNA), Brazil,
Process. No.~464898/2014-5.

\end{document}